\documentclass[a4paper,12pt]{article}
\pdfoutput=1

\usepackage[symbol]{footmisc}
\usepackage{jheppub}

\usepackage{amssymb,amsmath,bm}
\usepackage{color}
\usepackage{slashed}
\usepackage{stackengine}
\usepackage{hyperref}
\usepackage{relsize}
\usepackage{graphicx}
\usepackage[latin1]{inputenc}
\usepackage{amsfonts}
\usepackage{pifont}
\usepackage{amsthm}
\usepackage{booktabs}
\usepackage{array}
\usepackage{longtable}
\usepackage{lscape} 
\usepackage{rotating}
\usepackage{float}
\usepackage{multirow}
\usepackage{graphicx,rotating,amsmath,multirow}
\usepackage[table]{xcolor}
\usepackage{hyperref,slashed}
\usepackage{upgreek}

\usepackage{pdflscape} % Automatically rotate from Portrait to Landscape 

\usepackage{fontawesome} % GitHub symbol

\usepackage[detect-all]{siunitx} % For SI units
\usepackage{mathrsfs}            % For \mathscr

\graphicspath{{img/}}

\definecolor{rosso}{cmyk}{0,1,1,0.4}
\definecolor{rossos}{cmyk}{0,1,1,0.55}
\definecolor{rossoc}{cmyk}{0,1,1,0.2}
\definecolor{blu}{cmyk}{1,1,0,0.3}
\definecolor{blus}{cmyk}{1,1,0,0.6}
\definecolor{bluc}{cmyk}{1,1,0,0.1}
\definecolor{verde}{cmyk}{0.92,0,0.59,0.25}
\definecolor{verdec}{cmyk}{0.92,0,0.59,0.15}
\definecolor{verdes}{cmyk}{0.92,0,0.59,0.4}
\definecolor{Gray}{gray}{0.95}

\font\tenrsfs=rsfs10 at 12pt
\font\sevenrsfs=rsfs7
\font\fiversfs=rsfs5
\newfam\rsfsfam
\textfont\rsfsfam=\tenrsfs
\scriptfont\rsfsfam=\sevenrsfs
\scriptscriptfont\rsfsfam=\fiversfs

\usepackage[small,bf]{caption}
\usepackage{chngpage} % allows for temporary adjustment of side margins
\newcommand{\eps}{\varepsilon}
\newcommand{\vp}{H}
\newcommand{\tvp}{\widetilde{H}}

\newcommand{\vpj }{\mbox{${\vp^\dag i\,\raisebox{2mm}{\boldmath ${}^\leftrightarrow$}\hspace{-4mm} D_\mu\,\vp}$}}
\newcommand{\vpjt}{\mbox{${\vp^\dag i\,\raisebox{2mm}{\boldmath ${}^\leftrightarrow$}\hspace{-4mm} D_\mu^{\,I}\,\vp}$}}

\newcommand{\OO}{\ensuremath{\mathcal{O}}}

\newcommand{\sss}{\scriptscriptstyle}
\newcommand{\bpm}{\begin{pmatrix}}      
\newcommand{\epm}{\end{pmatrix}}

\newcommand{\Op}[1]{\OO_{\sss #1}}
\newcommand{\Opp}[2]{\OO_{\sss #1}^{\sss #2}}

\definecolor{lightgray}{rgb}{0.83, 0.83, 0.83}
\definecolor{lightpurp}{rgb}{0.901,0.796,0.882}

\newcommand{\lgc}{\cellcolor{lightgray} }

\definecolor{colorMag1}{RGB}{255, 255, 178} % Light yellow
\definecolor{colorMag2}{RGB}{254, 204, 92}  % Yellow
\definecolor{colorMag3}{RGB}{253, 141, 60}  % Orange
\definecolor{colorMag4}{RGB}{240, 59, 32}   % Red
\definecolor{colorMag5}{RGB}{189, 0, 38}    % Dark red

\newcommand{\bea}{\begin{eqnarray}}
\newcommand{\eea}{\end{eqnarray}}
\newcommand{\beq}{\begin{equation}}
\newcommand{\eeq}{\end{equation}}

\usepackage{tabularx}

\numberwithin{equation}{section}
\definecolor{niceRed}{rgb}{0.7,0.1,0.1}

\usepackage{tikz}
\usetikzlibrary{calc}

\newcommand{\makedictlink}[3]{\href{https://github.com/johngarg/linear-one-loop-dict-ref/blob/main/dict/#1/#2.txt}{\color{black}{#3}}}
\newcommand{\makeresultslink}[2]{\href{https://github.com/johngarg/linear-one-loop-dict-ref/blob/main/results/#1}{\color{black}{#2}}}

\newcommand{\midmidrule}{\arrayrulecolor{black!30}\midrule\arrayrulecolor{black}}

\title{\boldmath Linear Standard Model extensions in the SMEFT at one loop and Tera-$Z$}

\author[a,b,c]{John Gargalionis,}
\emailAdd{john.gargalionis@adelaide.edu.au}

\author[d]{J\'er\'emie Quevillon,}
\emailAdd{jeremie.quevillon@lapth.cnrs.fr}

\author[e]{Pham Ngoc Hoa Vuong,}
\emailAdd{hoa.vuong@desy.de}

\author[f]{Tevong You}
\emailAdd{Tevong.you@kcl.ac.uk}

\affiliation[a]{Departament de F\'isica Te\`orica, Universitat de Val\`encia, 46100 Burjassot, Spain}

\affiliation[b]{Instituto de F\'isica Corpuscular (CSIC-Universitat de Val\`encia), Parc Cient\'ific UV, C/Catedr\'atico Jos\'e Beltr\'an, 2, E-46980 Paterna, Spain}

\affiliation[c]{University of Adelaide, ARC Centre of Excellence for Dark Matter Particle Physics \& CSSM, Department of Physics, Adelaide SA 5000 Australia}

\affiliation[d]{
Laboratoire d'Annecy-le-Vieux de Physique Th\'{e}orique, CNRS -- USMB, BP 110 Annecy-le-Vieux, 74941 Annecy, France
}

\affiliation[e]{Deutsches Elektronen-Synchrotron DESY, Notkestr. 85, 22607 Hamburg, Germany}

\affiliation[f]{King's College London, Theoretical Particle Physics and Cosmology group, London,\\ United Kingdom}

\abstract{
Linear Standard Model (SM) extensions, defined as new particles that can couple linearly to SM fields, form a motivated and finite set of simplified models for exploring phenomenology Beyond the SM (BSM). Heavy BSM particles may be integrated out to obtain their low-energy effects in the SM Effective Field Theory (SMEFT) parametrised by the Wilson coefficients of higher-dimensional operators. We compute and map the dimension-6 SMEFT operator structure of all scalar and fermion linear SM extensions up to one-loop order, thus extending the existing tree-level dictionary of results. Explicit analytic matching expressions for the Wilson coefficients are provided as both Python and Mathematica code in a GitHub repository accessible through links embedded in our main table for each coefficient and within a Python package. We apply our map to highlight the sensitivity to heavy new physics of a $Z$-pole run at a future Tera-$Z$ factory; at one loop, with unit couplings, all linear SM extensions can be indirectly probed by electroweak precision measurements up to $\mathcal{O}(10)$ TeV.
}

\begin{document}
{
\rightline{KCL-PH-TH-2024-72}
\rightline{DESY-24-184}
\rightline{ADP-24-19/T1258}
}
\maketitle
\flushbottom
%\newpage

\section{Introduction}

There is no doubt as to the existence of new physics Beyond the Standard Model (BSM). However, the continued success of the Standard Model (SM) indicates that it must either be at higher energy scales or interact more elusively than experiments and observations can currently detect, or some combination of both. In the absence of unambiguous experimental clues or clear theoretical guidance for where BSM physics may first reveal itself, the continued exploration and experimental study of the zepto-scale must proceed more generally and with a more open mind.  

The Standard Model Effective Field Theory (SMEFT) is an appropriate framework for mapping the uncharted territory to explore~\cite{Weinberg:1979sa, Buchmuller:1985jz, Grzadkowski:2010es} (see e.g. Refs.~\cite{Brivio:2017vri, Falkowski:2023hsg, Isidori:2023pyp} for some recent reviews). It characterises the effects of heavy new physics in a general way, assuming only the SM degrees of freedom at low energies and the Higgs boson as part of a doublet in a linearly realised $\mathrm{SU}(2)_L \times \mathrm{U}(1)_Y$ low-energy theory of electroweak symmetry breaking.\footnote{See also e.g.~Refs.~\cite{Brivio:2016fzo, Cohen:2020xca, deBlas:2018tjm} for a non-linearly realised low-energy theory, the Higgs Effective Field Theory (HEFT), that relaxes this assumption.} Integrating out the heavy degrees of freedom generates contributions to the Wilson coefficients of the SMEFT, whose leading $CP$-even lepton-number-conserving effects first arise at mass dimension 6 in a series of increasingly suppressed higher-dimensional operators. A global fit to dimension-6 operators thus outlines the shape of excluded and allowed BSM physics models that are being probed or poorly constrained by measurements. Following the discovery of a Higgs boson~\cite{ATLAS:2012yve, CMS:2012qbp} compatible with SM expectation and no direct discovery of other BSM particles, such efforts have begun and are underway to characterise data in global fits to the SMEFT framework~\cite{Han:2004az, Pomarol:2013zra, Corbett:2012ja, Dumont:2013wma, Ellis:2014dva, Ellis:2014jta, Falkowski:2014tna, Efrati:2015eaa, Falkowski:2015krw, Buckley:2015lku, Berthier:2016tkq, deBlas:2016nqo, Falkowski:2017pss, Ellis:2018gqa, Aebischer:2018iyb, daSilvaAlmeida:2018iqo, Biekotter:2018ohn, Hartland:2019bjb, Falkowski:2019hvp, Aoude:2020dwv, Falkowski:2020pma, Ellis:2020unq, Bissmann:2020mfi, Ethier:2021bye, Grunwald:2023nli, Garosi:2023yxg, Allwicher:2023shc, Bartocci:2023nvp, Celada:2024mcf, deBlas:2022ofj}.

While the SMEFT provides a useful phenomenological parametrisation correlating disparate datasets, there are often too many coefficients to provide meaningful marginalised bounds without additional assumptions, and BSM models are expected to generate only a subset of coefficients in specific patterns. It is therefore crucial to interpret the results of constraints on the SMEFT in terms of simplified or realistic models, not only to reduce the dimensionality of the fit but also to provide information on the allowed ranges of masses and couplings that are being indirectly probed under different scenarios. This can also help to determine the validity of the EFT\@. The advantage of the SMEFT approach is to separate the indirect experimental signatures of heavy new physics from the model-specific interpretation, that can then be performed in a separate step without having to recalculate and redo the fit for every model.

A complementary approach to top-down model-building, especially in the current context of null results in BSM searches, is to use a more bottom-up phenomenological way of interpreting data in terms of simplified models (such a simplified model approach is used in dark matter searches~\cite{Abdallah:2014hon}, for example). One useful starting point is to consider linear SM extensions~\cite{Baldes:2011mh, deBlas:2014mba, deBlas:2017xtg, Herrero-Garcia:2019czj}, defined as BSM fields coupling linearly via marginal or relevant interactions to the SM. Their linear couplings enable them to be singly produced at colliders and have tree-level matchings to SMEFT operators. Ref.~\cite{deBlas:2017xtg} provided a complete classification of the finite set of linear SM extensions together with a dictionary of tree-level matching to dimension-6 operators. This streamlines the step of interpreting the SMEFT coefficients in terms of these UV models to a look-up table for leading-order tree-level effects. However, one-loop effects can also be phenomenologically relevant, and in some cases are indispensable for capturing the full picture. A partial one-loop dictionary for the subset of SMEFT operators at dimension six without tree-level completions was presented in Ref.~\cite{Guedes:2023azv}, and similar one-loop dictionaries have been produced for four-fermion operators~\cite{Cepedello:2022pyx, Cepedello:2023yao, Helo:2019yqp}, some dimension-eight operators~\cite{Cepedello:2024ogz}, and the dimension-five Weinberg operator~\cite{Bonnet:2012kz, Cai:2014kra, Gargalionis:2020xvt, Cai:2017jrq}. For recent advances in SMEFT at NLO, see e.g.~Refs.~\cite{Dawson:2019clf, Dawson:2022bxd,Bellafronte:2023amz,Degrande:2020evl}.

In this work we map out the one-loop structure of the SMEFT generated by integrating out the scalar and fermion linear SM extensions at one loop and provide a dictionary of explicit results for the Wilson coefficients. Recent advances in the methods and tools for one-loop matching computations~\cite{Henning:2014wua, Zhang:2016pja, Cohen:2020fcu, Drozd:2015rsp, Ellis:2016enq, Ellis:2017jns, Ellis:2020ivx, Larue:2023uyv, Li:2024ciy, Cohen:2020qvb, Carmona:2021xtq, Fuentes-Martin:2020udw, Fuentes-Martin:2022jrf, DasBakshi:2018vni, Banerjee:2023iiv, Chakrabortty:2023yke, DeAngelis:2023bmd,Chala:2024llp} enable such wide surveys of one-loop phenomenology to be carried out, with more systematic computations than ever before. In particular, the recently released {\tt Matchmaker}~\cite{Carmona:2021xtq} employs the traditional Feynman diagram approach while the nascent {\tt Matchete}~\cite{Fuentes-Martin:2020udw, Fuentes-Martin:2022jrf} automates functional methods that have seen recent developments improve their applicability for a more efficient approach to one-loop matching (see also {\tt Codex}~\cite{DasBakshi:2018vni}, a tool that implements some of the Universal One-Loop Effective Action (UOLEA) results~\cite{Drozd:2015rsp, Ellis:2016enq, Ellis:2017jns, Ellis:2020ivx, Larue:2023uyv, Li:2024ciy}). 

Fig.~\ref{fig:map} summarises our main result. Each row corresponds to a SMEFT operator and each column corresponds to a linear SM extension model. The entries are labelled T, L, or R depending on whether the operator is generated by the model via tree-level matching, one-loop matching, or one-loop renormalisation group running, respectively, and the colour shading is the numerical size of the operator coefficient setting all couplings to 1. Clicking on the hyperlinked entries in the table and on the particle names gives access to the full analytical expressions. For convenience, we also list in Appendix~\ref{sec:Table-couplings} and Table~\ref{tab:ewpo-op-couplings} the couplings for each model that are relevant for tree-level and one-loop matching to each operator, and describe the public code repositories for the results in Appendix~\ref{sec:repositories}.

Our map facilitates the interpretation of experimental constraints on SMEFT dimension-6 operator coefficients in terms of their implications for simplified models consisting of the scalar and fermion linear SM extensions considered here. Conversely, one could start with a simplified model of interest and immediately assess the relevant phenomenology in the SMEFT up to one loop order. The utility of this map is illustrated by considering the subset of dimension-6 operators relevant for electroweak precision measurements at the $Z$ pole whose ultra-high precision at a Tera-$Z$ factory would make it a powerful quantum probe of TeV-scale new physics~\cite{Allwicher:2023shc, Stefanek:2024kds, Allwicher:2024sso}. A limited number of simplified models can be constrained via their tree-level contributions to the SMEFT operators; on the other hand Fig.~\ref{fig:map} shows that all of them enter at one loop, and the relevant couplings can easily be read off for each operator. These are listed for convenience in Table~\ref{tab:ewpo-op-couplings} for the subset of electroweak precision operators, and in Appendix~\ref{sec:Table-couplings} for all other operators. While one-loop Renormalisation Group Evolution (RGE) effects from tree-level-generated operator mixings have previously been considered for some linear SM extensions in Refs.~\cite{Dawson:2020oco, Greljo:2023bdy, Celada:2024mcf} and for all the scalar, vector, and fermion models in Ref.~\cite{Allwicher:2024sso}, we incorporate here in full generality the finite contributions from the complete one-loop matching, including couplings that do not arise at tree level, which can also lead to important effects that enhance the $Z$-pole sensitivity.     

\begin{figure}[t]
  \centering
  \caption{Interactive (clickable) map of our one-loop dictionary~\href{https://github.com/johngarg/linear-one-loop-dict-ref}{\color{red}{\faGithub}}; see main text. \label{fig:map}}
  \resizebox{0.95\textwidth}{!}{
    \begin{tikzpicture}[scale=0.6]
      % Set the colour
      \foreach \y [count=\n] in {
        {, , , , , , , , 59, 59, 59, 55, 55, 53, 50, 50, 50, 44, 45, , , , , , , 55, 55, 51, 51, 51, 49, 49}, {, , , , , , , , , , , , , , , , , , , , , , , , , , , , , , , }, {, , , 80, 76, 72, 67, 67, , , , 74, 74, 66, , , , 62, 68, , , 76, 76, 72, 68, , , 70, 70, 70, 62, 62}, {, , , , , , , , , , , , , , , , , , , , , , , , , , , , , , , }, {-4, 39, 39, -1, 2, -5, 6, 2, 33, 33, 33, 20, 20, 22, 29, 29, 29, 18, -2, 11, 18, 18, 18, 50, 31, 10, 12, 8, 11, 10, 22, 25}, {48, 48, 40, 50, 42, 33, 48, 35, 54, 46, 38, 56, 34, 48, 50, 42, 35, 44, 36, 46, 39, 35, 33, 47, 41, 43, 52, 34, 37, 30, 54, 44}, {5, 43, 43, 25, 4, -2, 25, 23, 37, 37, 37, 29, 29, 33, 33, 33, 33, 29, 18, 14, 23, 24, 25, 26, 33, 18, 16, 9, 19, 17, 23, 25}, {, , , , , , , , , , , , , , , , , , , , , , , , , , , , , , , }, {31, 62, 54, 39, -4, -7, 25, 21, 68, 60, 53, 33, 33, 29, 64, 57, 49, 25, 13, 11, 22, 18, 20, 23, 28, 15, 14, 25, 13, 12, 18, 24}, {, , , , , , , , 45, 45, 45, 39, 39, 39, 36, 36, 36, 29, 27, , , , , , , 31, 31, 27, 31, 31, 32, 32}, {, , , , , , , , , , , , , , , , , , , , , , , , , , , , , , , }, {55, , , 57, 49, 46, 46, 46, , , , 50, 50, 45, , , , 41, 43, 53, 53, , , 49, 49, 46, 46, , , , 43, 43}, {48, , , 55, , 35, 42, 36, , , , 55, 44, 46, , , , 42, 40, 45, 41, 45, 45, 47, 45, 39, , 39, 45, 41, 49, 44}, {, , , , , , , , , , , , , , , , , , , , , , , , , , , , , , , }, {, , , , , , , , , , , , , , , , , , , , , , , , , , , , , , , }, {48, 68, 60, 35, 42, 34, 68, 56, 43, 28, 35, 36, 56, 65, 46, 30, 55, 63, 33, 45, 40, 39, 37, 47, 43, 37, 48, 4, 4, 33, 52, 40}, {42, 62, 29, 37, 36, 27, 62, 50, 39, 60, 34, 72, 28, 62, 64, 57, 49, 58, 54, 39, 34, 4, 4, 40, 36, 31, 42, 29, 29, 27, 45, 33}, {46, 28, 58, 36, 40, 3, 66, 54, 33, 64, 57, 37, 51, 27, 68, 60, 53, 62, 58, 8, 8, 37, 35, 10, 10, 35, 46, 33, 33, 31, 49, 37}, {53, 47, 90, 59, 52, 6, 71, 71, 38, , 91, 53, 54, 39, , , , 66, 72, 8, 8, 51, 51, 16, 16, 43, 43, 41, 45, 45, 44, 44}, {39, 72, 64, 37, 33, 25, 72, 60, 25, 70, 33, 81, 34, 22, 24, 66, 33, 22, 35, 49, 44, 43, 41, 51, 47, 8, 8, 39, 39, 33, 10, 10}, {41, , , 42, 41, 33, 71, 71, 31, 80, 33, 77, 41, 31, 36, 80, 33, 37, 40, 50, 50, 51, 51, 50, 50, 8, 8, 46, 45, 40, 16, 16}, {37, 64, 57, 33, 31, 23, 64, 52, 27, 63, 35, 75, 37, 22, 28, 59, 39, 23, 31, 41, 36, 35, 34, 43, 39, 33, 39, 4, 31, 4, 37, 36}, {55, , , 37, 49, , , , 30, , , , 60, 42, 29, , , 42, 35, , , , , , , 67, 67, 0, 65, 65, 59, 59}, {, , , 43, , , , , 34, 55, 62, 64, 62, 55, 34, 55, , 53, 42, , , , , , , 70, 73, 41, 63, , 75, 67}, {, , , 32, , , , , 26, 57, 65, 61, 65, 53, 34, 50, , 44, 38, , , , , , , 62, 72, 38, 60, , 73, 62}, {54, , , -0, 25, 25, 47, 47, 22, 44, 56, 49, 51, 36, 22, 44, , 42, 27, 41, 46, 47, 47, 52, 55, 41, 31, 23, 31, 40, 39, 35}, {, , , 54, , , , , 38, , , , 76, 53, 39, , , 54, 51, , , , , , , 69, 78, 49, 72, , 80, 74}, {, 75, 67, 22, , 68, 75, 62, 22, -0, 30, 24, 53, 52, 27, 8, 48, 43, 30, , 63, 67, 54, , 57, 46, 46, 38, 44, 41, 40, 40}, {, 60, 56, 78, , 53, , , 33, , 61, 69, 52, , , , , , , 78, 74, 78, 63, 80, 76, , , , , , , }, {57, 51, 51, -0, 28, 27, 50, 50, 23, , 53, 52, 23, 50, , , , 57, 42, 48, 33, 33, 33, 37, 41, 43, 43, 36, 44, 43, 50, 50}, {, 67, , 72, , 61, , , 40, , , 73, 47, 62, , , , , , 72, 81, , 75, 77, 83, , , , , , , }, {, 64, 31, 20, , 58, 64, 52, 19, 24, 4, 32, 35, 64, 30, 26, 51, 60, 35, , 53, 50, 46, , 46, 51, 59, 39, 38, 41, 53, 45}, {, 62, 6, 23, , 56, 62, 50, 20, 60, 24, 72, 19, 62, 64, 57, 49, 58, 54, , 50, 36, 52, , 44, 49, 57, 60, 42, 38, 50, 42}, {, 60, 27, 20, , 54, 60, 48, 3, 59, 14, 71, 22, 60, 35, 55, 22, 57, 39, , 49, 36, 37, , 42, 47, 55, 42, 40, 45, 49, 41}, {, 31, 60, 24, , 28, 68, 56, 21, 28, 35, 4, 33, 38, 34, 30, 55, 64, 39, 51, 47, 60, 48, 53, 54, 55, 63, 43, 42, 45, 57, 49}, {100, 24, 29, 4, 94, 22, 62, 50, 17, 60, 34, 33, 20, 31, 64, 57, 49, 58, 54, 45, 41, 48, 43, 46, 48, 49, 56, 60, 42, 38, 50, 42}, {93, , , -0, 87, 79, , , 8, , 80, 81, 53, 73, 53, , , , 59, 87, 83, 83, 80, 82, 84, 78, 76, 65, 76, 76, 78, 77}, {70, , , -0, 63, 55, , , 2, , , , 4, , 76, , , , 83, 63, 59, 59, 57, 59, 61, 53, 54, 49, 53, 52, 54, 55}, {, , , 37, , , , , 10, , , , 12, , , , , , , , , , , , , , , , , , , }, {, 0, 62, 27, 84, 13, 68, 57, 22, 68, 60, 22, 23, 13, 72, 64, 57, 64, 62, 31, 44, 62, 50, 43, 58, 57, 64, 65, 50, 46, 56, 50}, {, 35, 64, 38, , 32, 72, 60, 5, 71, 63, 43, 32, 6, 31, 67, 59, 28, 41, 42, 42, 64, 52, 44, 44, 46, 46, 71, 52, 48, 48, 49}, {, 47, , 24, 80, 41, 71, 71, 3, , , 53, 28, 4, 36, , , 34, 53, 44, 44, 72, 72, 53, 53, 55, 55, 66, 66, 66, 64, 64}, {, 27, 57, 26, , 24, 64, 52, 23, 63, 24, 35, 4, 37, 39, 59, 26, 60, 43, 40, 42, 57, 44, 42, 42, 51, 59, 46, 44, 49, 53, 45}, {100, 75, 67, 10, 94, 69, 75, 62, 15, 34, 42, 42, 38, 32, 18, 36, 61, 30, 9, 90, 63, 67, 54, 92, 57, 48, 49, 38, 38, 51, 50, 51}, {90, , , -1, 84, 76, , , 11, 29, 38, 34, 39, 10, 19, 22, 39, 16, 11, 80, 80, 80, 80, 82, 82, 66, 67, 55, 62, 32, 32, 32}, {, 68, 35, 36, , 62, 68, 56, 13, 67, 40, 79, 5, 26, 27, 63, 55, 24, 37, , 57, 39, 38, , 50, 42, 42, 67, 48, 45, 44, 45}, {, 83, 75, 35, , 76, 83, 70, 6, 66, 65, 62, 29, 2, 2, 57, 56, 4, 30, , 71, 75, 62, , 64, 45, 40, 50, 50, 49, 41, 50}, {, , , 22, 84, 80, 75, 75, 3, 60, 60, 56, 45, 4, 8, 51, 51, 20, 37, , , 76, 76, 72, 68, 46, 46, 44, 44, 44, 44, 44}, {54, 71, 63, 10, 47, 39, 71, 58, 20, 69, 34, 81, 38, 33, 18, 65, 34, 27, 9, 44, 43, 43, 42, 45, 44, 35, 47, 31, 37, 41, 40, 36}, {44, , , -1, 37, 29, , , 7, 48, 27, 44, 21, 12, 16, 39, 22, 16, 11, 33, 33, 33, 33, 35, 35, 27, 29, 23, 25, 25, 25, 26}, {67, , , -0, 61, 53, , , 6, , , , 84, 59, 9, , , , 28, 57, 57, 57, 57, 58, 58, 51, 51, 39, 49, 49, 51, 51}, {, , , 12, , , , , -8, , , , , 45, -3, , , , 1, , , , , , , , , , , , , }, {, , , 37, , , , , 46, , 28, , 34, 31, 37, , 28, 30, 50, , , , , , , 49, 54, 49, , 38, 46, 50}, {, , , 30, , , , , 35, , 34, , 36, 29, 26, , 27, 21, 29, , , , , , , 48, 39, 45, , 36, 39, 50}, {28, , , -0, 2, 1, 24, 23, 18, , 21, 39, 24, 16, 19, , 21, 12, 12, 17, 23, 24, 24, 28, 31, 7, 18, 11, 18, 8, 11, 15}, {, , , 41, , , , , 36, , , , 53, 30, 37, , , 31, 47, , , , , , , 55, 46, 72, , 49, 51, 57}, {, 67, 59, 27, , 60, 67, 54, 6, 26, 21, 34, 37, 67, 6, 29, 29, 63, 27, , 55, 59, 46, , 49, 53, 61, 48, 38, 51, 55, 47}, {, , , 10, , , , , -0, 29, 12, 34, 34, 42, 5, 22, 22, 33, 25, , , , , , , 36, 36, 27, 61, 61, 30, 30}, {, 67, 59, 22, , 60, 67, 54, 20, 57, -0, 54, 26, 52, 27, 48, 8, 43, 29, , 55, 59, 46, , 49, 45, 46, 41, 40, 34, 40, 39},
      } {
        % heatmap tiles
        \foreach \x [count=\m] in \y {
          % Define the hyperlink values
          \node[fill=white!\x!verdes, minimum size=6mm] at (\m,-\n) {};
        }
      }

      % Set the label
      \foreach \y [count=\n] in {
        {, , , , , , , , L, L, L, L, L, L, L, L, L, L, L, , , , , , , L, L, L, L, L, L, L}, {, , , , , , , , , , , , , , , , , , , , , , , , , , , , , , , }, {, , , L, L, L, L, L, , , , L, L, L, , , , L, L, , , L, L, L, L, , , L, L, L, L, L}, {, , , , , , , , , , , , , , , , , , , , , , , , , , , , , , , }, {TLR, L, L, TLR, TLR, TLR, TLR, TLR, L, L, L, L, L, L, L, L, L, L, L, LR, LR, LR, LR, LR, LR, LR, LR, LR, LR, LR, LR, LR}, {L, L, L, L, L, L, L, L, L, L, L, L, L, L, L, L, L, L, L, L, L, L, L, L, L, L, L, L, L, L, L, L}, {TLR, L, L, L, TLR, TLR, L, L, L, L, L, L, L, L, L, L, L, L, L, LR, LR, LR, LR, LR, LR, LR, LR, LR, LR, LR, LR, LR}, {, , , , , , , , , , , , , , , , , , , , , , , , , , , , , , , }, {L, L, L, L, TLR, TLR, L, L, L, L, L, L, L, L, L, L, L, L, L, LR, LR, LR, LR, LR, LR, LR, LR, LR, LR, LR, LR, LR}, {, , , , , , , , L, L, L, L, L, L, L, L, L, L, L, , , , , , , L, L, L, L, L, L, L}, {, , , , , , , , , , , , , , , , , , , , , , , , , , , , , , , }, {L, , , L, L, L, L, L, , , , L, L, L, , , , L, L, L, L, , , L, L, L, L, , , , L, L}, {L, , , L, , L, L, L, , , , L, L, L, , , , L, L, L, L, L, L, L, L, L, , L, L, L, L, L}, {, , , , , , , , , , , , , , , , , , , , , , , , , , , , , , , }, {, , , , , , , , , , , , , , , , , , , , , , , , , , , , , , , }, {LR, L, L, LR, LR, LR, L, L, LR, LR, LR, LR, L, L, LR, LR, L, L, LR, L, L, L, L, L, L, LR, LR, TLR, TLR, L, LR, LR}, {LR, L, LR, LR, LR, LR, L, L, LR, L, LR, L, LR, L, L, L, L, L, L, LR, LR, TLR, TLR, LR, LR, L, L, LR, L, L, L, L}, {LR, LR, L, LR, LR, LR, L, L, LR, L, L, LR, LR, LR, L, L, L, L, L, TLR, TLR, LR, LR, TLR, TLR, L, L, LR, L, L, L, L}, {LR, LR, L, L, LR, LR, L, L, LR, , L, L, L, LR, , , , L, L, TLR, TLR, L, L, TLR, TLR, L, L, L, L, L, L, L}, {LR, L, L, LR, LR, LR, L, L, LR, L, L, L, LR, LR, LR, L, L, LR, LR, L, L, L, L, L, L, TLR, TLR, LR, LR, LR, TLR, TLR}, {LR, , , L, LR, LR, L, L, LR, L, L, L, L, LR, LR, L, L, LR, L, L, L, L, L, L, L, TLR, TLR, L, L, L, TLR, TLR}, {LR, L, L, LR, LR, LR, L, L, LR, L, LR, L, LR, L, LR, L, LR, L, LR, L, L, L, L, L, L, LR, LR, TLR, L, TLR, LR, LR}, {LR, , , L, LR, , , , LR, , , , L, L, LR, , , L, L, , , , , , , L, L, TLR, LR, LR, L, L}, {, , , LR, , , , , LR, L, L, L, L, L, LR, L, , L, LR, , , , , , , L, L, L, L, , L, L}, {, , , LR, , , , , LR, L, L, L, L, L, LR, L, , L, LR, , , , , , , L, L, L, L, , L, L}, {LR, , , TLR, TLR, TLR, L, L, LR, L, L, L, L, L, LR, L, , L, LR, LR, LR, LR, LR, LR, LR, LR, TLR, TLR, TLR, LR, TLR, TLR}, {, , , LR, , , , , LR, , , , L, L, LR, , , L, LR, , , , , , , L, L, L, L, , L, L}, {, L, L, LR, , L, L, L, LR, TLR, LR, LR, L, L, LR, TLR, L, L, LR, , L, L, L, , L, L, L, LR, LR, L, L, L}, {, L, L, L, , L, , , LR, , L, L, LR, , , , , , , L, L, L, L, L, L, , , , , , , }, {LR, L, L, TLR, TLR, TLR, L, L, LR, , L, L, LR, L, , , , L, L, LR, TLR, TLR, TLR, TLR, TLR, LR, LR, LR, LR, LR, LR, LR}, {, L, , L, , L, , , LR, , , L, LR, L, , , , , , L, L, , L, L, L, , , , , , , }, {, L, L, LR, , L, L, L, LR, L, TLR, LR, LR, L, L, L, L, L, L, , L, LR, LR, , L, L, L, LR, LR, L, L, L}, {, L, TLR, LR, , L, L, L, LR, L, LR, L, LR, L, L, L, L, L, L, , L, LR, LR, , L, L, L, L, L, L, L, L}, {, L, L, LR, , L, L, L, TLR, L, LR, L, LR, L, L, L, L, L, L, , L, LR, LR, , L, L, L, LR, L, LR, L, L}, {, L, L, LR, , L, L, L, LR, L, LR, TLR, L, LR, L, L, L, L, L, LR, LR, L, L, LR, LR, L, L, LR, LR, L, L, L}, {L, LR, LR, TLR, L, LR, L, L, LR, L, L, L, LR, L, L, L, L, L, L, LR, LR, LR, LR, LR, LR, L, L, L, L, L, L, L}, {L, , , TLR, L, L, , , LR, , LR, LR, LR, LR, LR, , , , LR, L, L, L, L, L, L, L, L, L, L, L, L, L}, {L, , , TLR, L, L, , , TLR, , , , TLR, , LR, , , , LR, L, L, L, L, L, L, L, L, L, L, L, L, L}, {, , , LR, , , , , TLR, , , , TLR, , , , , , , , , , , , , , , , , , , }, {, TLR, L, LR, L, TLR, L, L, LR, L, L, LR, LR, LR, L, L, L, L, L, LR, LR, L, L, LR, LR, L, L, L, L, L, L, L}, {, L, L, LR, , L, L, L, TLR, L, L, LR, LR, TLR, L, L, L, LR, LR, LR, LR, L, L, LR, LR, LR, LR, L, L, L, LR, LR}, {, L, , LR, L, L, L, L, TLR, , , L, LR, TLR, L, , , LR, L, LR, LR, L, L, LR, LR, LR, LR, L, L, L, LR, LR}, {, L, L, LR, , L, L, L, LR, L, L, L, TLR, LR, L, L, L, L, L, LR, LR, L, L, LR, LR, L, L, LR, L, LR, L, L}, {L, L, L, TLR, L, L, L, L, LR, LR, L, L, L, LR, LR, LR, L, LR, TLR, L, L, L, L, L, L, LR, LR, LR, LR, L, LR, LR}, {L, , , TLR, L, L, , , LR, LR, L, L, L, LR, LR, LR, L, LR, TLR, L, L, L, L, L, L, L, L, L, L, L, L, L}, {, L, L, LR, , L, L, L, LR, L, LR, L, TLR, LR, L, L, L, LR, LR, , L, LR, LR, , L, LR, LR, L, L, L, LR, LR}, {, L, L, LR, , L, L, L, TLR, L, L, L, LR, TLR, TLR, L, L, TLR, LR, , L, L, L, , L, LR, LR, L, L, L, LR, LR}, {, , , LR, L, L, L, L, TLR, L, L, L, L, TLR, TLR, L, L, TLR, LR, , , L, L, L, L, LR, LR, L, L, L, LR, LR}, {L, L, L, TLR, L, L, L, L, LR, L, LR, L, LR, LR, LR, L, LR, LR, TLR, L, L, L, L, L, L, LR, LR, LR, L, LR, LR, LR}, {L, , , TLR, L, L, , , LR, L, LR, L, LR, LR, LR, L, LR, LR, TLR, L, L, L, L, L, L, L, L, L, L, L, L, L}, {L, , , TLR, L, L, , , TLR, , , , LR, L, TLR, , , , LR, L, L, L, L, L, L, L, L, L, L, L, L, L}, {, , , LR, , , , , TLR, , , , , L, TLR, , , , TLR, , , , , , , , , , , , , }, {, , , LR, , , , , LR, , L, , LR, L, LR, , L, L, LR, , , , , , , L, L, L, , L, L, L}, {, , , LR, , , , , LR, , L, , L, L, LR, , L, L, LR, , , , , , , L, L, L, , L, L, L}, {LR, , , TLR, TLR, TLR, L, L, LR, , L, L, LR, L, LR, , L, L, LR, LR, LR, LR, LR, LR, LR, TLR, LR, TLR, LR, TLR, TLR, TLR}, {, , , LR, , , , , LR, , , , LR, L, LR, , , L, LR, , , , , , , L, L, L, , L, L, L}, {, L, L, LR, , L, L, L, TLR, L, LR, L, L, L, TLR, L, L, L, LR, , L, L, L, , L, L, L, LR, LR, LR, L, L}, {, , , LR, , , , , TLR, L, L, L, L, L, TLR, L, L, L, LR, , , , , , , L, L, LR, L, L, L, L}, {, L, L, LR, , L, L, L, LR, L, TLR, L, LR, L, LR, L, TLR, L, LR, , L, L, L, , L, L, L, LR, L, LR, L, L},
      } {
        % heatmap tiles
        \foreach \x [count=\m] in \y {
          % Define the hyperlink values
          \node[font=\tiny] at (\m,-\n) {\makedictlink{\m}{\n}{\x}};
        }
      }

      % row labels
      \foreach \a [count=\i] in {$\mathcal{O}_{G}$,
        $\mathcal{O}_{\tilde{G}}$,
        $\mathcal{O}_{W}$,
        $\mathcal{O}_{\tilde{W}}$,
        $\mathcal{O}_{H}$,
        $\mathcal{O}_{HB}$,
        $\mathcal{O}_{H D^2}$,
        $\mathcal{O}_{H\tilde{B}}$,
        $\mathcal{O}_{HD}$,
        $\mathcal{O}_{HG}$,
        $\mathcal{O}_{H\tilde{G}}$,
        $\mathcal{O}_{HW}$,
        $\mathcal{O}_{HWB}$,
        $\mathcal{O}_{HW\tilde{B}}$,
        $\mathcal{O}_{H\tilde{W}}$,
        $\mathcal{O}_{Hd}$,
        $\mathcal{O}_{He}$,
        $\mathcal{O}_{Hl}^{(1)}$,
        $\mathcal{O}_{Hl}^{(3)}$,
        $\mathcal{O}_{Hq}^{(1)}$,
        $\mathcal{O}_{Hq}^{(3)}$,
        $\mathcal{O}_{Hu}$,
        $\mathcal{O}_{Hud}$,
        $\mathcal{O}_{dB}$,
        $\mathcal{O}_{dG}$,
        $\mathcal{O}_{dH}$,
        $\mathcal{O}_{dW}$,
        $\mathcal{O}_{dd}$,
        $\mathcal{O}_{eB}$,
        $\mathcal{O}_{eH}$,
        $\mathcal{O}_{eW}$,
        $\mathcal{O}_{ed}$,
        $\mathcal{O}_{ee}$,
        $\mathcal{O}_{eu}$,
        $\mathcal{O}_{ld}$,
        $\mathcal{O}_{le}$,
        $\mathcal{O}_{ledq}$,
        $\mathcal{O}_{lequ}^{(1)}$,
        $\mathcal{O}_{lequ}^{(3)}$,
        $\mathcal{O}_{ll}$,
        $\mathcal{O}_{lq}^{(1)}$,
        $\mathcal{O}_{lq}^{(3)}$,
        $\mathcal{O}_{lu}$,
        $\mathcal{O}_{qd}^{(1)}$,
        $\mathcal{O}_{qd}^{(8)}$,
        $\mathcal{O}_{qe}$,
        $\mathcal{O}_{qq}^{(1)}$,
        $\mathcal{O}_{qq}^{(3)}$,
        $\mathcal{O}_{qu}^{(1)}$,
        $\mathcal{O}_{qu}^{(8)}$,
        $\mathcal{O}_{quqd}^{(1)}$,
        $\mathcal{O}_{quqd}^{(8)}$,
        $\mathcal{O}_{uB}$,
        $\mathcal{O}_{uG}$,
        $\mathcal{O}_{uH}$,
        $\mathcal{O}_{uW}$,
        $\mathcal{O}_{ud}^{(1)}$,
        $\mathcal{O}_{ud}^{(8)}$,
        $\mathcal{O}_{uu}$
      } {
        \node[minimum size=6mm,font=\small,text width=20mm] at (0,-\i) {\a};
      }

      % Column labels
      \foreach \a [count=\i] in {
        $S$,
        $S_1$,
        $S_2$,
        $\varphi$,
        $\Xi$,
        $\Xi_1$,
        $\Theta_1$,
        $\Theta_3$,
        $\omega_1$,
        $\omega_2$,
        $\omega_4$,
        $\Pi_1$,
        $\Pi_7$,
        $\zeta$,
        $\Omega_1$,
        $\Omega_2$,
        $\Omega_4$,
        $\Upsilon$,
        $\Phi$,
        $N$,
        $E$,
        $\Delta_1$,
        $\Delta_3$,
        $\Sigma$,
        $\Sigma_1$,
        $U$,
        $D$,
        $Q_1$,
        $Q_5$,
        $Q_7$,
        $T_1$,
        $T_2$,
      } {
        \node[minimum size=6mm,font=\small] at (\i,0) {\makeresultslink{\i}{\a}};
      }

      % Add a color bar
      \foreach \y in {0, 1, ..., 100} {
        \node[fill=white!\y!verdes, minimum size=8mm] at (36,-\y/5-20) {};
      }
      \node[anchor=west, font=\small] at (36.8, 1-20) {$\geq 10^0$};
      \node[anchor=west, font=\small] at (36.8, -24.125) {$10^{-2}$};
      \node[anchor=west, font=\small] at (36.8, -29.25) {$10^{-4}$};
      \node[anchor=west, font=\small] at (36.8, -34.375) {$10^{-6}$};
      \node[anchor=west, font=\small] at (36.8, -20-20+0.5) {$\leq 10^{-8}$};

      % Label the color bar
      \node[anchor=south, rotate=90] at (35.2, -10-20) {$|C_i|~\mathrm{TeV}^{-2}$};
    \end{tikzpicture}
  }
\end{figure}

This paper is organised as follows. In Section~\ref{sec: Lin-SM-extensions}, we outline the general classification of linear SM extensions that form the set of simplified models considered in this work, including the Lagrangian relevant for loop-level matching, as adapted from Ref.~\cite{deBlas:2017xtg} that we supplemented by adding further terms relevant at the one-loop level. Section~\ref{sec:One-loop map of SMEFT} details the mapping of their tree-level and one-loop structure in the SMEFT, summarised in Fig.~\ref{fig:map}. Finally, an application of our map is provided in Section~\ref{sec:teraZ} by highlighting the sensitivity to the linear SM extensions of a $Z$-pole run at a future Tera-$Z$ factory where the one-loop contributions play a crucial role in obtaining complete coverage. We conclude in Section~\ref{sec:Conclusion}. The public code repositories are described in Appendix~\ref{sec:repositories}. Appendix~\ref{sec:Table-couplings} and Table~\ref{tab:ewpo-op-couplings} contain tables of the relevant model couplings necessary for generating each operator.

\section{Linear Standard Model extensions}
\label{sec: Lin-SM-extensions}

\begin{table}[ht!]
\begin{minipage}[t]{0.48\linewidth}
    \centering
    \scalebox{0.95}{
	\begin{tabular}{|c|c|c|}
		\hline
		Name & Irrep & Examples  \\
		\hline
		\hline
		$\mathcal{S}$ & $(1,1)_0$ & Singlet scalar~\cite{Haisch:2020ahr} \\
		\hline 
		$\mathcal{S}_1$ & $(1,1)_1$ & Zee model~\cite{zee1980389, Zee:1985id} \\ 
		\hline
		$\mathcal{S}_2$ & $(1,1)_2$ & Zee--Babu model~\cite{Zee:1985id, Babu:1988ki} \\ 
		\hline
		$\varphi$ & $\left(1,2\right)_{\frac 12}$  & 2HDM~\cite{Branco:2011iw} \\
		\hline
		$\Xi$ &	$\left(1,3\right)_0$ & Georgi-Machacek~\cite{Chanowitz:1985ug, Georgi:1985nv} \\
		\hline
		$\Xi_1$ & $\left(1,3\right)_1$ & Type-II seesaw~\cite{Magg:1980ut,
  PhysRevD.22.2227, LAZARIDES1981287, Wetterich:1981bx, PhysRevD.23.165} \\
		\hline
		$\Theta_1$ & $\left(1,4\right)_{\frac 12}$ & Quartet~\cite{Cirelli:2005uq,Durieux:2022hbu,Anamiati:2018cuq} \\
		\hline
		$\Theta_3$ & $\left(1,4\right)_{\frac 32}$ & Quartet~\cite{Cirelli:2005uq,Babu:2009aq,Durieux:2022hbu} \\
		\hline
		${\omega}_{1}$  & $\left(3,1\right)_{-\frac 13}$ & Leptoquark $S_1$~\cite{Dorsner:2016wpm}  \\
		\hline
		${\omega}_{2}$ & $\left(3,1\right)_{\frac 23}$ & Leptoquark $\bar{S}_1$~\cite{Dorsner:2016wpm}  \\
		\hline
		${\omega}_{4}$ & $\left(3,1\right)_{-\frac 43}$ & Leptoquark $\tilde{S}_1$~\cite{Dorsner:2016wpm}  \\
		\hline
		$\Pi_1$ & $\left(3,2\right)_{\frac 16}$ & Leptoquark $\tilde{R}_2$~\cite{Dorsner:2016wpm}  \\
		\hline
		$\Pi_7$ & $\left(3,2\right)_{\frac 76}$ & Leptoquark $R_2$~\cite{Dorsner:2016wpm}  \\
		\hline
		$\zeta$ & $\left(3,3\right)_{-\frac 13}$ & Leptoquark $S_3$~\cite{Dorsner:2016wpm}  \\
		\hline
		$\Omega_{1}$ & $\left(6,1\right)_{\frac 13}$ & Diquark~\cite{Giudice:2011ak} \\
		\hline
		$\Omega_{2}$ & $\left(6,1\right)_{-\frac 23}$ & Diquark~\cite{Marciano:1980zf,Giudice:2011ak,Englert:2024nlj} \\
		\hline
		$\Omega_{4}$ & $\left(6,1\right)_{\frac 43}$ & Diquark~\cite{Marciano:1980zf,Giudice:2011ak} \\
		\hline
		$\Upsilon$ & $\left(6,3\right)_{\frac 13}$ & Diquark~\cite{Marciano:1980zf,Giudice:2011ak} \\
		\hline
		$\Phi$ & $\left(8,2\right)_{\frac 12}$ & Manohar--Wise~\cite{Manohar:2006ga} \\
		\hline
	\end{tabular}
    }
\end{minipage}
    \hfill
\begin{minipage}[t]{0.48\linewidth}
    \centering
    \scalebox{0.95}{
	\begin{tabular}{|c|c|c|}
		\hline
		Name & Irrep & Examples \\
		\hline
		\hline
		$N$  & $\left(1, 1\right)_0$ & Type-I seesaw~\cite{MINKOWSKI1977421, Yanagida:1979as, GellMann:1980vs, PhysRevLett.44.912, Glashow:1979nm} \\
		\hline
		$E$ & $\left(1, 1\right)_{-1}$ & Singlet VLL~\cite{Falkowski:2013jya, Altmannshofer:2013zba} \\
		\hline
		$\Delta_1$ & $\left(1, 2\right)_{-\frac{1}{2}}$ & Doublet VLL~\cite{Falkowski:2013jya, Altmannshofer:2013zba} \\
		\hline
		$\Delta_3$ & $\left(1, 2\right)_{-\frac{3}{2}}$ & Doublet VLL~\cite{Erdelyi:2025axy} \\
		\hline
		$\Sigma$ & $\left(1, 3\right)_0$ & Type-III seesaw~\cite{Foot:1988aq} \\
		\hline
		$\Sigma_1$ & $\left(1, 3\right)_{-1}$ & Triplet VLL~\cite{Cirelli:2005uq,Babu:2009aq} \\
		\hline
		$U$ & $\left(3, 1\right)_{\frac{2}{3}}$ & Singlet VLQ, $T$~\cite{Aguilar-Saavedra:2009xmz} \\
		\hline
		$D$ & $\left(3, 1\right)_{-\frac{1}{3}}$ & Singlet VLQ, $B$~\cite{Aguilar-Saavedra:2009xmz} \\
		\hline
		$Q_1$ & $\left(3, 2\right)_{\frac{1}{6}}$ & Doublet VLQ, $(TB)$~\cite{Aguilar-Saavedra:2009xmz} \\
		\hline
		$Q_5$ & $\left(3, 2\right)_{-\frac{5}{6}}$ & Doublet VLQ, $(BY)$~\cite{Aguilar-Saavedra:2009xmz} \\
		\hline
		$Q_7$ & $\left(3, 2\right)_{\frac{7}{6}}$ & Doublet VLQ, $(XT)$~\cite{Aguilar-Saavedra:2009xmz} \\
		\hline
		$T_1$ & $\left(3, 3\right)_{-\frac{1}{3}}$ &  Triplet VLQ $(TBY)$~\cite{Aguilar-Saavedra:2013qpa,Buchkremer:2013bha} \\
		\hline
		$T_2$ & $\left(3, 3\right)_{\frac{2}{3}}$ &  Triplet VLQ $(XTB)$~\cite{Aguilar-Saavedra:2013qpa,Buchkremer:2013bha} \\
		\hline
        \end{tabular}
    }
\end{minipage}
\caption{\it Table of scalar fields (left) and fermion fields (right) that can couple linearly and renormalisably to the SM listed according to their irreducible representations under ${\mathrm{SU}(3)}_{C} \times {\mathrm{SU}(2)}_{L} \times {\mathrm{U}(1)}_{Y}$. Some common names and alternative notations in use are given, with examples of some references in the literature where these models have been studied outside of the context of SUSY.
\label{tab:scalars-fermions}
}
\end{table}

Amongst all possible extensions of the SM, those coupling linearly to SM fields form a motivated and finite subset for a bottom-up, simplified-model approach. Their linear couplings allow for tree-level matching to higher-dimensional operators when integrated out. They can also be singly produced at colliders. One may then investigate systematically the phenomenological implications of indirect and direct measurements for this class of models, to leading order, by enumerating the linear SM extensions and computing their SMEFT Wilson coefficients at tree-level. This programme was carried out in Ref.~\cite{deBlas:2017xtg} using a Lagrangian sufficient for tree-level matching. Below we summarise the scalar and fermion field content of the linear SM extensions shown in Table~\ref{tab:scalars-fermions} and extend the Lagrangian presented in Ref.~\cite{deBlas:2017xtg} to accommodate loop-level matching using the results of Refs.~\cite{DasBakshi:2021xbl, Naskar:2022rpg}. The vector field extension requires further UV-dependent considerations for finite one-loop matching that we leave to future work.\footnote{
Integrating out heavy vector fields at one loop is currently not supported by the existing automated tools such as {\tt Matchmaker}~\cite{Carmona:2021xtq} and {\tt Matchete}~\cite{Fuentes-Martin:2020udw, Fuentes-Martin:2022jrf}. At one loop, the vector extensions will be divergent without specifying the mass generation mechanism and the origin of the longitudinal modes. We do not have a complete description of this in our simplified model approach; introducing these longitudinal modes would go beyond one of the assumptions of our framework, namely that we are working only with single-multiplet extensions of the SM.  
}

\subsection{Scalar and fermion multiplets}

Table~\ref{tab:scalars-fermions} lists the scalar and fermion fields separately, classified by their irreducible representations under the SM gauge group: ${\mathrm{SU}(3)}_{C} \times {\mathrm{SU}(2)}_{L} \times {\mathrm{U}(1)}_{Y}$, with $Q=I_3+Y$. Some common names for the fields are given, as are examples of references in which their phenomenology is studied outside the context of supersymmetry. In supersymmetric theories the linear couplings would violate $R$-parity and their phenomenology could depart significantly from the simplified model approach.  

Our objects of study are simplified models that extend the SM by only one\footnote{Here we are counting vector-like Dirac fermions as one multiplet.} of these exotic multiplets. To accommodate this, our approach is to introduce only a limited set of Lorentz representations: real and complex scalars, vector-like Dirac fermions and Majorana fermions. This ensures that the simplified models we write down are UV complete, in the sense that they are consistent theories with good high-energy behaviour, and allows us to ensure a separation of scales for which the SMEFT description is valid, since the masses of these fields are free parameters in the simplified models. The introduction of chiral fermions necessitates a number of model-building complications associated with mass generation and anomaly cancellation, and exotic Proca fields need to be interpreted in the context of some larger UV framework to be treated consistently at loop level. In addition, we assume a single generation for all exotic fields.

\subsection{Full Lagrangian for one-loop matching}
\label{sec:full-lagrangian}

In Table~\ref{tab:full-lag} we present the Lagrangian we use to perform the one-loop matching using \texttt{Matchmaker}. We remind the reader that each UV model consists of the SM particle content extended by only one of the linear SM extensions at a time. For this reason, mixed terms that are relevant for matching up to one-loop are not listed below. Additionally, we only include terms relevant for one-loop matching, as laid out by Refs.~\cite{DasBakshi:2021xbl, Naskar:2022rpg}. Thus, even for a specific multiplet the Lagrangian terms listed are not the full Lagrangian of the model, but are sufficient for our purposes.

In some cases, Fermi--Dirac statistics impose that some Yukawa couplings are antisymmetric matrices in flavour. For example, the couplings ${[y_{\mathcal{S}_1}]}_{pq}$ appearing in the term ${[y_{\mathcal{S}_1}]}_{pq} \mathcal{S}_{1}^\dagger \bar{L}_{ip} {L^{\mathsf c}_{jq}} {\epsilon}^{i j}$ are necessarily antisymmetric under the exchange $p \leftrightarrow q$. In such cases, we have explicitly shown the antisymmetry by putting square brackets around the flavour indices.

{\small
\begin{longtable}{ll}
\caption{\label{tab:full-lag}The table shows the terms in the interaction Lagrangian, $\mathcal{L}_\mathrm{int} = -\sum_M \mathcal{L}_M$, sufficient for one-loop matching onto the SMEFT grouped by exotic multiplet $M$. Couplings with a hat denote those that do not enter in tree-level matching. To simplify notation, chirality subscripts have been suppressed on the fermion fields: explicitly, $e$, $d$ and $u$ stand for $e_R$, $d_R$ and $u_R$, while $Q$ and $L$ are $Q_L$ and $L_L$. Bold typesetting is used for tensors with indices enumerating components of representations larger than the fundamental for $\mathrm{SU}(2)$ and $\mathrm{SU}(3)$. Adjoint indices $I,J,K$ are often used for $\mathrm{SU}(2)$. Indices whose contraction is unique are sometimes omitted and the associated fields are enclosed in parentheses. Flavour indices are taken from $p,q,r,s$ and are antisymmetric under exchange if enclosed in square brackets.} \\
\toprule
$M$ & Terms \\
\midrule
\endfirsthead

\toprule
$M$ & Terms \\
\midrule
\endhead

$\mathcal{S}$ & $\kappa_{\mathcal{S}} \mathcal{S} (H^\dagger H) +  \lambda_{\mathcal{S}} \mathcal{S} \mathcal{S} (H^\dagger H) + \kappa_{\mathcal{S}^3} \mathcal{S} \mathcal{S} \mathcal{S}$ \\

\midmidrule
$\mathcal{S}_1$ & ${[y_{\mathcal{S}_1}]}_{[pq]} \mathcal{S}_{1}^\dagger \bar{L}_{ip} {L^{\mathsf c}_{jq}} {\epsilon}^{i j} + \hat{\lambda}_{\mathcal{S}_1} (H^\dagger H) \mathcal{S}_{1}^\dagger \mathcal{S}_{1}$ \\
\midmidrule
$\mathcal{S}_2$ & ${[y_{\mathcal{S}_2}]}_{p q} \mathcal{S}_{2}^\dagger \bar{e}_{p} e_{q}^{\mathsf c} + \hat{\lambda}_{\mathcal{S}_2} (H^\dagger H) \mathcal{S}_{2}^\dagger \mathcal{S}_{2}$ \\
\midmidrule
\multirow{2}{*}{$\varphi$} & ${[y_{\varphi e}]}_{pq} \varphi^\dagger_i \bar{e}_{p} L_{q}^i + {[y_{\varphi d}]}_{pq} \varphi^\dagger_{i} \bar{d}_{p} Q^{i}_{q} + {[y_{\varphi u}]}_{pq} \varphi^\dagger_{i} \bar{Q}_{jp} u_{q} \epsilon^{i j} + \lambda_{\varphi} (\varphi^\dagger H) (H^\dagger H)$ \\
 & $+ \hat{\lambda}_{\varphi} (H^\dagger H) (\varphi^\dagger \varphi) + \hat{\lambda}^{\prime}_{\varphi} (H^\dagger \varphi) (\varphi^\dagger H)$ \\
\midmidrule
$\Xi$ & $\kappa_{\Xi} H^\dagger (\boldsymbol{\Xi} \cdot \boldsymbol{\sigma}) H + \lambda_{\Xi} (\boldsymbol{\Xi} \cdot \boldsymbol{\Xi}) (H^\dagger H)$ \\

\midmidrule
\multirow{2}{*}{$\Xi_1$} & $\frac{1}{2} \lambda_{\Xi_1} (\mathbf{\Xi_{1}^{\dagger}} \cdot \mathbf{\Xi_{1}}) (H^\dagger H) + \frac{1}{2} {\lambda_{\Xi_1}^\prime} \Xi_{1}^{\dagger I} \Xi_{1}^{J} (H^\dagger {\sigma}^{K} H) f_{I J K} + {[y_{\Xi_1}]}_{p q} (\mathbf{\Xi_{1}^{\dagger}} \cdot \boldsymbol{\sigma})^{i}_{\ j} \bar{L}_{ip} L^{\mathsf{c}}_{kq} {\epsilon}^{j k}$ \\
& $+~ \kappa_{\Xi_1} (\mathbf{\Xi_{1}^{\dagger}} \cdot \boldsymbol{\sigma})_{\ k}^{j} H^{i} H^{k} \epsilon_{i j}$ \\
\midmidrule

\multirow{2}{*}{$\Theta_1$} & $\lambda_{\Theta_1} H^\dagger_{i} H^{j} H^\dagger_{k} {\epsilon}^{kl} ({\mathbf{C_{2224}}} \cdot \mathbf{\Theta_{1}})_{j l}^{i}+ \hat{\lambda}_{\Theta_1} (H^\dagger H) (\boldsymbol{\Theta}_{1}^\dagger \cdot \boldsymbol{\Theta}_{1}) $ \\
& $+ \hat{\lambda}^{\prime}_{\Theta_1} (\boldsymbol{\Theta_{1}}^\dagger \cdot \mathbf{T_{4}} \cdot \boldsymbol{\Theta_{1}})^{I} (H^\dagger {\sigma} H)^{I} + \hat{\lambda}^{\prime\prime}_{\Theta_1} (\boldsymbol{\Theta_{1}} \cdot \mathbf{C_{2244}} \cdot \boldsymbol{\Theta_{1}})^{i j} H^\dagger_{i} H^\dagger_{j}$ \\
\midmidrule

\multirow{2}{*}{$\Theta_3$} & $\lambda_{\Theta_3} H^\dagger_{i} H^\dagger_{j}  H^\dagger_{k} (\mathbf{C_{2224}} \cdot \mathbf{\Theta_{3}})_{l m}^{i} \epsilon^{j l} \epsilon^{k m} + \hat{\lambda}_{\Theta_3} (H^\dagger H) (\boldsymbol{\Theta_{3}}^\dagger \cdot \boldsymbol{\Theta_{3}})$ \\
& $ + \hat{\lambda}^{\prime}_{\Theta_3} (\boldsymbol{\Theta_{3}}^\dagger \cdot \mathbf{T_{4}} \cdot \boldsymbol{\Theta_{3}})^{I} (H^\dagger {\sigma} H)^{I}$ \\
\midmidrule

\multirow{2}{*}{$\omega_{1}$} & ${[y_{q\ell \omega_1}]}_{p q} \omega_{1}^\dagger \bar{Q}^{\mathsf{c}\, i}_{p} L^{j}_{q} {\epsilon}_{i j} + {[y_{qq \omega_1}]}_{p q} \omega_{1\, a}^\dagger \bar{Q}_{b i p} {Q^{\mathsf{c}}}_{c j q} {\epsilon}^{i j} \epsilon^{abc} + {[y_{e u \omega_1}]}_{pq} \omega_{1}^\dagger {\bar{e}^{\mathsf{c}}}_{p} u_{q}$ \\
& $+ {[y_{d u \omega_1}]}_{p q} \omega_{1\, a}^\dagger \bar{d}_{b p} u^{\mathsf{c}}_{c q} \epsilon^{a b c} + \hat{\lambda}_{\omega_1} (H^\dagger H) (\omega_{1}^\dagger \omega_{1})$ \\

\midmidrule
$\omega_{2}$ & ${[y_{\omega_2}]}_{[p q]} \omega_{2\, a}^\dagger \bar{d}_{b p} d^{\mathsf{c}}_{c q} {\epsilon}^{abc} + \hat{\lambda}_{\omega_2} (H^\dagger H) (\omega_{2}^\dagger \omega_{2})$ \\

\midmidrule
$\omega_{4}$ & ${[y_{e d \omega_4}]}_{p q} \omega_{4}^\dagger \bar{e}^{\mathsf{c}}_{p} d_{q} + {[y_{u u \omega_4}]}_{[pq]} \omega_{4\, a}^\dagger \bar{u}_{bp} u^{\mathsf{c}}_{c q} \epsilon^{abc} + \hat{\lambda}_{\omega_4} (H^\dagger H) (\omega_{4}^\dagger \omega_{4})$ \\

\midmidrule
$\Pi_{1}$ & ${[y_{\Pi_1}]}_{pq} \Pi_{1\, i}^\dagger \bar{L}_{jp} d_{q} \epsilon^{i j} + \hat{\lambda}_{\Pi_1} (H^\dagger H) (\Pi_{1}^\dagger \Pi_{1}) + \hat{\lambda}^{\prime}_{\Pi_1} (\Pi_{1}^\dagger H)_{a} (H^\dagger \Pi_{1})^{a}$ \\

\midmidrule
\multirow{2}{*}{$\Pi_{7}$} & ${[y_{\ell u \Pi_7}]}_{pq} \Pi_{7\, i}^\dagger \bar{L}_{j p} u_{q} \epsilon^{i j}  + {[y_{e q \Pi_7}]}_{p q} \Pi_{7\, i}^\dagger \bar{e}_{p} Q^{i}_{q} + \hat{\lambda}_{\Pi_7} (H^\dagger H) (\Pi_{7}^\dagger \Pi_{7})$ \\
& $+ \hat{\lambda}^{\prime}_{\Pi_7} (\Pi_{7}^\dagger H)_{a} (H^\dagger \Pi_{7})^{a}$ \\

\midmidrule
\multirow{2}{*}{$\zeta$} & ${[y_{q\ell \zeta}]}_{pq} (\boldsymbol{\zeta}^\dagger \cdot \boldsymbol{\sigma})_{j}^{k} \bar{Q}^{\mathsf{c}\, i}_{p} L^{j}_{q} \epsilon_{i k} + {[y_{qq \zeta}]}_{[pq]} (\boldsymbol{\zeta}^\dagger_{a} \cdot \boldsymbol{\sigma})_{k}^{i} \bar{Q}_{b i p} Q^{\mathsf{c}}_{c j q} \epsilon^{k j} \epsilon^{abc} + \hat{\lambda}_{\zeta} (H^\dagger H) (\boldsymbol{\zeta}^\dagger \cdot \boldsymbol{\zeta})$ \\
& $+ \hat{\lambda}^{\prime}_{\zeta} (\zeta^{\dagger I} \zeta^{J}) (H^\dagger \sigma^{K} H) f_{I J K}$ \\

\midmidrule
$\Omega_{1}$ & ${[y_{u d \Omega_1}]}_{pq} (\mathbf{K} \cdot \boldsymbol{\Omega_{1}}^\dagger)_{ab} \bar{u}^{\mathsf{c}\, a}_{p} d^{b}_{q} + {[y_{q q \Omega_1}]}_{[p q]} (\mathbf{K} \cdot \boldsymbol{\Omega_{1}}^\dagger)_{ab} \bar{Q}^{\mathsf{c}\, ci}_p Q^{b j}_q \epsilon_{i j} + \hat{\lambda}_{\Omega_1} (H^\dagger H) (\boldsymbol{\Omega_{1}}^\dagger \cdot \boldsymbol{\Omega_{1}})$ \\

\midmidrule
$\Omega_{2}$ & ${[y_{\Omega_2}]}_{pq} (\mathbf{K} \cdot \boldsymbol{\Omega_{2}}^\dagger)_{ab} {\bar{d}^{\mathsf{c}\, a}}_{p} d_{q}^{b} + \hat{\lambda}_{\Omega_2} (H^\dagger H) (\boldsymbol{\Omega_{2}}^\dagger \cdot \boldsymbol{\Omega_{2}})$ \\

\midmidrule
$\Omega_{4}$ & ${[y_{\Omega_4}]}_{pq} (\mathbf{K} \cdot \boldsymbol{\Omega_{4}}^\dagger)_{ab} \bar{u}^{\mathsf{c}\, a}_{p} u^{b}_q + \hat{\lambda}_{\Omega_4} (H^\dagger H) (\boldsymbol{\Omega_{4}}^\dagger \cdot \boldsymbol{\Omega_{4}})$ \\

\midmidrule
$\Upsilon$ & ${[y_{\Upsilon}]}_{pq} (\mathbf{K} \cdot \boldsymbol{\Upsilon}^{\dagger})^{I}_{ab} (\bar{Q}^{\mathsf{c}}_{p} \epsilon \sigma Q_{q})^{Iab} + \hat{\lambda}_{\Upsilon} (H^\dagger H) (\boldsymbol{\Upsilon}^\dagger \cdot \boldsymbol{\Upsilon}) + \hat{\lambda}^{\prime}_{\Upsilon} (\boldsymbol{\Upsilon}^{\dagger} \cdot \boldsymbol{\Upsilon})^{IJ} (H^\dagger \sigma H)^{K} f_{I J K}$ \\

\midmidrule
\multirow{2}{*}{$\Phi$} & $\frac{1}{2} {[y_{q u \Phi}]}_{pq} \bar{Q}_{a j p} (\boldsymbol{\lambda} \cdot \boldsymbol{\Phi}^\dagger_i)_{b}^{a} u^{b}_{q} \epsilon^{i j} + \frac{1}{2} {[y_{q d \Phi}]}_{pq} \bar{d}_{ap} (\boldsymbol{\lambda} \cdot \boldsymbol{\Phi}^\dagger_{i})^{a}_{b} Q^{b i}_{q} + \hat{\lambda}_{\Phi} (H^\dagger H) (\boldsymbol{\Phi}^\dagger \cdot \boldsymbol{\Phi})$ \\
& $+\hat{\lambda}^{\prime}_{\Phi} \mathrm{Tr}[(\boldsymbol{\Phi}^\dagger \cdot \boldsymbol{\lambda} H) (H^\dagger \boldsymbol{\Phi} \cdot \boldsymbol{\lambda})]  + \hat{\lambda}^{\prime\prime}_{\Phi} \mathrm{Tr}[(H^\dagger \boldsymbol{\Phi} \cdot \boldsymbol{\lambda}) (H^\dagger \boldsymbol{\Phi} \cdot \boldsymbol{\lambda})]$ \\

\midmidrule
$N$ & ${[\lambda_N]}_{p} \bar{N} L^{i}_{p} H^{j} \epsilon_{i j} $ \\

\midmidrule
$E$ & ${[\lambda_E]}_{p} \bar{E} L^{i}_{p} H^\dagger_{i}$ \\

\midmidrule
$\Delta_1$ & ${[\lambda_{\Delta_1}]}_{p} \bar{\Delta}_{1\, i} e_p H^{i}$ \\

\midmidrule
$\Delta_3$ & ${[\lambda_{\Delta_3}]}_{p} \bar{\Delta}_{3\, i} e_{p} H^\dagger_{j} \epsilon^{i j}$ \\

\midmidrule
$\Sigma$ & $\frac{1}{2} {[\lambda_{\Sigma}]}_{p} (\bar{\boldsymbol{\Sigma}} \cdot \boldsymbol{\sigma})_{i}^{j} L^{i}_{p} H^{k} \epsilon_{k j} $ \\

\midmidrule
$\Sigma_1$ & $\frac{1}{2} {[\lambda_{\Sigma_1}]}_{p} (\bar{\boldsymbol{\Sigma}}_{\mathbf{1}} \cdot \boldsymbol{\sigma})_{i}^{j} L^{i}_{p}  H^\dagger_{j}$ \\

\midmidrule
$U$ & ${[\lambda_U]}_{p} \bar{U}_{a} Q^{a i}_{p} H^{j} \epsilon_{i j}$ \\

\midmidrule
$D$ & ${[\lambda_D]}_{p} \bar{D}_{a} Q^{a i}_{p} H^\dagger_{i}$ \\

\midmidrule
$Q_1$ & ${[\lambda_{u Q_1}]}_{p} \bar{Q}_{1\, a i} u^{a}_{p} H^\dagger_{i} \epsilon^{i j} + {[\lambda_{d Q_1}]}_{p} \bar{Q}_{1\,a i} d^{a}_{p} H^{i}$ \\

\midmidrule
$Q_5$ & ${[\lambda_{Q_5}]}_{p} \bar{Q}_{5\, a i} d^{a}_{p} H^\dagger_{j} \epsilon^{i j}$ \\

\midmidrule
$Q_7$ & ${[\lambda_{Q_7}]}_{p} \bar{Q}_{7\, a i} u^{a}_{p} H^{i}$ \\

\midmidrule
$T_1$ & $\frac{1}{2} {[\lambda_{T_1}]}_{p} (\bar{\mathbf{T}}_{\mathbf{1}\,a} \cdot \boldsymbol{\sigma})_{i}^{j} Q^{a i}_{p} H^\dagger_{j} $ \\

\midmidrule
$T_2$ & $\frac{1}{2} {[\lambda_{T_2}]}_{p} (\bar{\mathbf{T}}_{\mathbf{2}\,a} \cdot \boldsymbol{\sigma})_{i}^{j} Q^{a i}_{p} H^{k} {\epsilon}_{k j} $ \\

\bottomrule
\end{longtable}
}

The Lagrangian includes a number of group-theory factors and Clebsch-Gordan coefficients. For the contraction of three isospin doublets and an $\mathrm{SU}(2)$ quartet, we use
\begin{equation}
  [C_{2224}]^{i}_{jkR} \equiv [\sigma^{I}]^{i}_{~j} C^{QI}_{k} \epsilon_{QR} \; ,
\end{equation}
where the indices $P,Q,R$ are quartet indices, and $I$ is an adjoint index. The tensors $C^{QI}_k$ and $\epsilon_{QR}$ are defined in Ref.~\cite{deBlas:2017xtg}. For contracting two quartets and a triplet, we use the $T_\mathbf{4}^I$ matrices introduced in Eq.~\eqref{eq:T4-matrices}, and therefore two doublets and two quartets can be formed into a singlet with
\begin{equation}
  [C_{2244}]^{ij}_{RP} \equiv \frac{4}{\sqrt{15}} [\sigma^{I}]^{i}_{~k} \epsilon^{kj} [T_{\mathbf{4}}]^{I}_{QP} \epsilon_{RQ} \; ,
\end{equation}
as shown explicitly in the penultimate line of Eq.~\eqref{eq:new-terms}. We remind the reader that we use $f_{IJK}$ as defined in Ref.~\cite{deBlas:2017xtg}.

We use $[\lambda^{A}]^{a}_{b}$ for the components of the Gell--Mann matrices, and the $\mathbf{K}$ matrices as defined in Ref.~\cite{Han:2009ya} to contract two colour triplets and a colour sextet. Where the contraction of indices is unique, we sometimes omit them for notational clarity and wrap the expression within parentheses. For example, $(\boldsymbol{\Theta_{1}} \cdot \mathbf{C_{2244}} \cdot \boldsymbol{\Theta_{1}})^{i j}$ stands for $\Theta_1^P [C_{2244}]^{ij}_{PQ} \Theta_1^Q$, with a sum over the quartet indices $P$ and $Q$.

We have corrected some typos from Ref.~\cite{deBlas:2017xtg} and included the interactions necessary for generating one-loop diagrams. These terms were omitted from the Lagrangian of Ref.~\cite{deBlas:2017xtg}, which was sufficient only for tree-level matching at dimension 6. These additional terms contain only interactions between scalars, and we have derived them using the formulae presented in Ref.~\cite{DasBakshi:2021xbl}. Where possible we have kept the notation of Ref.~\cite{deBlas:2017xtg}, and we distinguish the new terms that we have added that do not enter at tree level but are necessary at one loop with a hat on the coefficients.

The additions necessary for the scalar part of the Lagrangian include modifications to the cubic and quartic interactions between exotic scalars and the SM Higgs. Adding only one exotic scalar multiplet at a time, Table 2 of Ref.~\cite{DasBakshi:2021xbl} suggests the only additional terms that need to be added are
\begin{equation}
  \label{eq:new-terms}
  \begin{aligned}
    \Delta \mathcal{L}_{\text{scalar}} &= \sum_S \hat{\lambda}_{S} (H^\dagger H) (S^\dagger S) + \hat{\lambda}_{\varphi}^\prime (H^\dagger \varphi) (\varphi^\dagger H) + \sum_{i \in \{1,3\}} \hat{\lambda}_{\Theta_i}^\prime (\boldsymbol{\Theta}_i^\dagger \cdot \mathbf{T}_{\mathbf{4}}^I \cdot \boldsymbol{\Theta}_i) (H^\dagger \sigma^I H) \\
    &\quad + \sum_{i \in\{1,7\}}\hat{\lambda}^\prime_{\Pi_i} (\Pi_i^\dagger H) (H^\dagger \Pi_i)  + \hat{\lambda}^\prime_{\Phi} \mathrm{Tr}[(\mathbf{\Phi}^\dagger \cdot \boldsymbol{\lambda} H) (H^\dagger \mathbf{\Phi} \cdot \boldsymbol{\lambda})] \\
    &\quad +  \sum_{S \in \{\zeta, \Upsilon\}}\hat{\lambda}_{S}^\prime f_{IJK}(S^{I\dagger} S^J) (H^\dagger \sigma^K H) + \left\{
    \hat{\lambda}^{\prime\prime}_{\Theta_1} \frac{4}{\sqrt{15}} (\Theta_1^P \epsilon_{PQ} [T_{\mathbf{4}}^I]_{~R}^{Q} \Theta_1^R) (H^\dagger \sigma^I \tilde{H}) \right. \\ &\quad \left. +  \hat{\lambda}^{\prime\prime}_{\Phi}\mathrm{Tr}[(H^\dagger \mathbf{\Phi} \cdot \boldsymbol{\lambda})(H^\dagger \mathbf{\Phi} \cdot \boldsymbol{\lambda})] + \text{h.c.} \vphantom{\frac{8}{3\sqrt{5}}} \right\}
    \ .
  \end{aligned}
\end{equation}
Here the first sum is taken over complex scalars excluding the multiplet $\Xi_1$, for which the appropriate terms are already included in Ref.~\cite{deBlas:2017xtg}. We note that only the scalars $\mathcal{S}$ and $\Xi$ are real and all others are complex; thus the sum over $S$ in the first term should be understood as a sum over all scalars excluding $\mathcal{S}$ and $\Xi$. Traces are taken over colour indices, and $\boldsymbol{\lambda}$ is a vector of the Gell-Mann matrices. The $\epsilon_{PQ}$ and $f_{IJK}$ are defined in Ref.~\cite{deBlas:2017xtg}, and the $\mathbf{T}_{\mathbf{4}}^I$ matrices are the generators of $\mathrm{SU}(2)$ in the isospin-$3/2$ representation:
\begin{equation}
\label{eq:T4-matrices}
\begin{split}
  \mathbf{T}_{\mathbf{4}}^1 &= \frac{1}{2}
  \begin{pmatrix}
  0 & \sqrt{3} & 0 & 0 \\
  \sqrt{3} & 0 & 2 & 0 \\
  0 & 2 & 0 & \sqrt{3} \\
  0 & 0 & \sqrt{3} & 0 \\
  \end{pmatrix},~
  \mathbf{T}_{\mathbf{4}}^2 = \frac{i}{2}
  \begin{pmatrix}
  0 & -\sqrt{3} & 0 & 0 \\
  \sqrt{3} & 0 & -2 & 0 \\
  0 & 2 & 0 & -\sqrt{3} \\
  0 & 0 & \sqrt{3} & 0 \\
  \end{pmatrix}, \\
  \mathbf{T}_{\mathbf{4}}^3 &= \frac{1}{2} \mathrm{diag}(3,1,-1,-3) \ .
  \end{split}
\end{equation}
Note that here we are taking $I,J,K$ as $\mathrm{SU}(2)$ adjoint indices, and $P,Q,R$ as quartet indices.

\section{One-loop SMEFT map of linear SM extensions}
\label{sec:One-loop map of SMEFT}

\subsection{The Standard Model Effective Field Theory}
\label{sec: SMEFT}

In the SMEFT framework, the usual SM Lagrangian with renormalisable terms up to mass dimension four is supplemented by higher-dimensional operators (see e.g. Refs.~\cite{Brivio:2017vri, Falkowski:2023hsg, Isidori:2023pyp} for a review and Refs.~\cite{Ellis:2020unq, Bissmann:2020mfi, Ethier:2021bye, Grunwald:2023nli, Garosi:2023yxg, Allwicher:2023shc, Bartocci:2023nvp, Celada:2024mcf, deBlas:2022ofj} for some recent global fits). The leading lepton-number-conserving, $CP$-even effects generically arise at dimension six, with higher operator dimensions suppressed by correspondingly larger inverse powers of some EFT cut-off scale. The SMEFT Lagrangian in this approximation can then be written as
\begin{equation}
	\mathcal{L}_{\text{SMEFT}} \supset \mathcal{L}_{\text{SM}} + \sum_{i=1}^{2499} \frac{c_i}{\Lambda^2}\mathcal{O}_i \, ,
\end{equation}
where $c_i$ are the dimensionless Wilson coefficients of the $B$-conserving, dimension-6 operators $\mathcal{O}_i$ and $\Lambda$ is the heavy ultra-violet (UV) scale where new BSM degrees of freedom generating the Wilson coefficients enter to replace the SMEFT low-energy description. Under the most general flavour assumptions, there are 2499 such operators in a non-redundant basis~\cite{Alonso:2013hga}. We shall work in the so-called Warsaw basis convention of Ref.~\cite{Grzadkowski:2010es}. The operators are reproduced here for convenience in Table~\ref{tab:warsawbosonic} and Table~\ref{tab:warsawfourfermion}. Operators shaded grey denote those that can only be generated at loop level.

\begin{table}[t!]
	{\small
		\centering
		%\hspace{-5mm}
		\renewcommand{\arraystretch}{1.0}
		\begin{tabular}{||c|c||c|c||c|c|} 
			\hline
			\lgc $\Op{G}$                 & $f^{ABC} G_\mu^{A\nu} G_\nu^{B\rho} G_\rho^{C\mu} $ &  $\Op{\vp}$               & $(\vp^\dag\vp)^3$ &
			\lgc $\Op{eW}$               & $(\bar l_p \sigma^{\mu\nu} e_r) \tau^I \vp W_{\mu\nu}^I$ \\
			\hline
			\lgc $\Op{W}$                 & $\eps^{IJK} W_\mu^{I\nu} W_\nu^{J\rho} W_\rho^{K\mu}$ &    
			$\Op{\vp\Box}$        & $(\vp^\dag \vp)\raisebox{-.5mm}{$\Box$}(\vp^\dag \vp)$ &
			\lgc $\Op{eB}$               &  $(\bar l_p \sigma^{\mu\nu} e_r) \vp B_{\mu\nu}$  \\
			\hline
			\lgc $\Op{\vp G}$            & $\vp^\dag \vp\, G^A_{\mu\nu} G^{A\mu\nu}$ & 
			$\Op{\vp D}$          & $\left(\vp^\dag D^\mu\vp\right)^\star \left(\vp^\dag D_\mu\vp\right)$ &
			\lgc $\Op{uG}$               &  $(\bar q_p \sigma^{\mu\nu} T^A u_r) \tvp\, G_{\mu\nu}^A$    \\
			\hline
			\lgc $\Op{\vp W}$            & $\vp^\dag \vp\, W^I_{\mu\nu} W^{I\mu\nu}$ &  
			 $\Opp{\vp l}{(1)}$      & $(\vpj)(\bar l_p \gamma^\mu l_r)$	&
			 \lgc $\Op{uW}$               &  $(\bar q_p \sigma^{\mu\nu} u_r) \tau^I \tvp\, W_{\mu\nu}^I$  \\
			\hline
			\lgc $\Op{\vp B}$            & $ \vp^\dag \vp\, B_{\mu\nu} B^{\mu\nu}$ &
			$\Opp{\vp l}{(3)}$      & $(\vpjt)(\bar l_p \tau^I \gamma^\mu l_r)$	&
			\lgc $\Op{uB}$               &  $(\bar q_p \sigma^{\mu\nu} u_r) \tvp\, B_{\mu\nu}$ \\
			\hline
			\lgc $\Op{\vp WB}$           & $ \vp^\dag \tau^I \vp\, W^I_{\mu\nu} B^{\mu\nu}$ &
			$\Op{\vp e}$            & $(\vpj)(\bar e_p \gamma^\mu e_r)$	&
			\lgc $\Op{dG}$               & $(\bar q_p \sigma^{\mu\nu} T^A d_r) \vp\, G_{\mu\nu}^A$ \\
			\hline
			 $\Op{e\vp}$           &  $(\vp^\dag \vp)(\bar l_p e_r \vp)$ &
			 $\Opp{\vp q}{(1)}$      & $(\vpj)(\bar q_p \gamma^\mu q_r)$	&
			 \lgc $\Op{dW}$               &  $(\bar q_p \sigma^{\mu\nu} d_r) \tau^I \vp\, W_{\mu\nu}^I$	\\
			\hline
			$\Op{u\vp}$           &  $(\vp^\dag \vp)(\bar q_p u_r \tvp)$ &
			$\Opp{\vp q}{(3)}$      & $(\vpjt)(\bar q_p \tau^I \gamma^\mu q_r)$	&
			\lgc $\Op{dB}$               &  $(\bar q_p \sigma^{\mu\nu} d_r) \vp\, B_{\mu\nu}$	\\
			\hline
			$\Op{d\vp}$           &  $(\vp^\dag \vp)(\bar q_p d_r \vp)$  &
			$\Op{\vp u}$            & $(\vpj)(\bar u_p \gamma^\mu u_r)$ &
			&	\\
			\hline
			 $\Op{\vp u d}$          & $i(\tvp^\dag D_\mu \vp)(\bar u_p \gamma^\mu d_r)$ &
			 $\Op{\vp d}$            & $(\vpj)(\bar d_p \gamma^\mu d_r)$& 
			 & \\
			\hline
		\end{tabular}
	}
	%\end{adjustwidth}
\caption{\it $CP$-even dimension-6 operators involving bosonic fields in the Warsaw basis. Grey shading indicates operators that can only be generated at loop level, while unshaded operators can receive tree-level as well as one-loop contributions. \label{tab:warsawbosonic}}
\end{table}

\begin{table}[t!]
	%\begin{adjustwidth}{-0.28in}{-0.28in}% adjust the L and R margins 
	{\small
		\centering
		%\hspace{-5mm}
		\renewcommand{\arraystretch}{1.0}
		\begin{tabular}{||c|c||c|c||c|c|} 
			\hline
			$\Op{ll}$               & $(\bar l_p \gamma_\mu l_r)(\bar l_s \gamma^\mu l_t)$ & 
			$\Op{eu}$                      & $(\bar e_p \gamma_\mu e_r)(\bar u_s \gamma^\mu u_t)$	& 
			$\Opp{qu}{(1)}$         & $(\bar q_p \gamma_\mu q_r)(\bar u_s \gamma^\mu u_t)$  \\
			\hline
			$\Opp{qq}{(1)}$  & $(\bar q_p \gamma_\mu q_r)(\bar q_s \gamma^\mu q_t)$ &
			$\Op{ed}$                      & $(\bar e_p \gamma_\mu e_r)(\bar d_s\gamma^\mu d_t)$ &
			$\Opp{qu}{(8)}$         & $(\bar q_p \gamma_\mu T^A q_r)(\bar u_s \gamma^\mu T^A u_t)$  \\
			\hline
			$\Opp{qq}{(3)}$  & $(\bar q_p \gamma_\mu \tau^I q_r)(\bar q_s \gamma^\mu \tau^I q_t)$ &
			$\Opp{ud}{(1)}$                & $(\bar u_p \gamma_\mu u_r)(\bar d_s \gamma^\mu d_t)$ &
			$\Opp{qd}{(1)}$ & $(\bar q_p \gamma_\mu q_r)(\bar d_s \gamma^\mu d_t)$  \\
			\hline
			$\Opp{lq}{(1)}$                & $(\bar l_p \gamma_\mu l_r)(\bar q_s \gamma^\mu q_t)$ &
			$\Opp{ud}{(8)}$                & $(\bar u_p \gamma_\mu T^A u_r)(\bar d_s \gamma^\mu T^A d_t)$ &
			$\Opp{qd}{(8)}$ & $(\bar q_p \gamma_\mu T^A q_r)(\bar d_s \gamma^\mu T^A d_t)$ \\
			\hline
			$\Opp{lq}{(3)}$                & $(\bar l_p \gamma_\mu \tau^I l_r)(\bar q_s \gamma^\mu \tau^I q_t)$ &
			$\Op{le}$               & $(\bar l_p \gamma_\mu l_r)(\bar e_s \gamma^\mu e_t)$ &
			$\Op{ledq}$  & $(\bar l_p^j e_r)(\bar d_s q_t^j)$  \\
			\hline
			$\Op{ee}$               & $(\bar e_p \gamma_\mu e_r)(\bar e_s \gamma^\mu e_t)$ &
			$\Op{lu}$               & $(\bar l_p \gamma_\mu l_r)(\bar u_s \gamma^\mu u_t)$  &
			$\Opp{quqd}{(1)}$ &  $(\bar q_p^j u_r) \eps_{jk} (\bar q_s^k d_t)$ \\
			\hline
			$\Op{uu}$        & $(\bar u_p \gamma_\mu u_r)(\bar u_s \gamma^\mu u_t)$ &
			$\Op{ld}$               & $(\bar l_p \gamma_\mu l_r)(\bar d_s \gamma^\mu d_t)$ &
			$\Opp{quqd}{(8)}$ & $(\bar q_p^j T^A u_r) \eps_{jk} (\bar q_s^k T^A d_t)$ \\
			\hline
			$\Op{dd}$        & $(\bar d_p \gamma_\mu d_r)(\bar d_s \gamma^\mu d_t)$ &
			$\Op{qe}$               & $(\bar q_p \gamma_\mu q_r)(\bar e_s \gamma^\mu e_t)$ &
			$\Opp{lequ}{(1)}$ &  $(\bar l_p^j e_r) \eps_{jk} (\bar q_s^k u_t)$ \\
			\hline
			 & & 
			 & & 
			$\Opp{lequ}{(3)}$ & $(\bar l_p^j \sigma_{\mu\nu} e_r) \eps_{jk} (\bar q_s^k \sigma^{\mu\nu} u_t)$ \\
			\hline
		\end{tabular}
	}
\caption{\it Four-fermion $B$-conserving dimension-6 operators in the Warsaw basis.
\label{tab:warsawfourfermion}}
\end{table}

\subsection{Methodology}\label{sec:methods}

Our approach is to leverage the impressive recent advances in automating the process of one-loop matching by using \texttt{Matchmaker} to construct the one-loop UV/IR dictionary for the linear SM extensions. We have written some auxiliary code to streamline the procedure of generating the inputs to the programme and to connect the output with the growing suite of HEP tools within the Python ecosystem. Below we discuss these tools, how we use them, and how their use can be generalised.

The \texttt{Matchmaker} programme takes as input a user-defined \texttt{FeynRules}~\cite{Alloul:2013bka} file specifying the model's interactions and couplings, along with an additional file defining the EFT onto which the UV theory is matched. We take the EFT to be the baryon-number-preserving SMEFT at dimension 6, for which the \texttt{Matchmaker} authors have provided an example EFT file. To facilitate the process of generating the UV-model files, we created a custom tool that parses user-defined interaction terms, performs various consistency checks, and outputs the Lagrangian to \texttt{FeynRules} format and LaTeX. The code can be straightforwardly extended to accommodate output to other formats as well. These output \texttt{FeynRules} files are then used as inputs to the main one-loop-matching function of \texttt{Matchmaker}. Additional auxiliary files including flavour relations between UV parameters and the definition of Clebsch--Gordan coefficients are prepared by hand for use with \texttt{Matchmaker}. We note that the production of the input \texttt{FeynRules} files has now been completely automated in the tool \texttt{SOLD}~\cite{Guedes:2023azv} through the group-theory functionality of \texttt{GroupMath}~\cite{Fonseca:2020vke}.

% We consider extending the SM by only one of the linear SM extensions at a time. In each case, we try and choose the Lagrangian terms to match the normalisations and conventions of the existing tree-level dictionary~\cite{deBlas:2017xtg}, although in some cases we depart from these, \textit{e.g.} for colour sextet contractions with anti-triplets, for which we use the $\mathbf{K}$ matrices defined in Ref.~\cite{Han:2009ya}.

We consider extending the SM by only one of the linear SM extensions at a time. In each case, we choose the Lagrangian terms to match the normalisations and conventions of the existing tree-level dictionary~\cite{deBlas:2017xtg}.

The one-loop matching results produced by \texttt{Matchmaker} are written to a file called \texttt{MatchingResult.dat} containing a Wolfram-language replacement list mapping operator-coefficient names to Mathematica expressions representing the matching contributions to those operators. We have written a custom tool, \texttt{MatchMakerParser}~\href{https://github.com/johngarg/MatchMakerParser}{\color{red}{\faGithub}}, designed to parse the matching-result dictionary provided by \texttt{Matchmaker} and translate the expressions into a Python class, facilitating further analysis within the Python ecosystem. \texttt{MatchMakerParser} is implemented in Mathematica and requires only the path to the \texttt{MatchingResult.dat} file and a declaration of the free parameters in the model. Within each class, Mathematica symbols are interpreted as object attributes and the matching expressions for individual operator coefficients are accessible through eponymous methods, whose arguments are the zero-indexed flavour indices on the SMEFT coefficient. We provide an easy way to access these Python classes through the \texttt{lsme}~\href{https://github.com/johngarg/lsme}{\color{red}{\faGithub}} package. This bridge between \texttt{Matchmaker} and the Python ecosystem allows for convenient use of our dictionary with tools such as, for example, \texttt{Flavio}~\cite{Straub:2018kue}, \texttt{Smelli}~\cite{Stangl:2020lbh}, SMEFiT~\cite{Giani:2023gfq}, and \texttt{Fitmaker}~\cite{Ellis:2020unq}, thus streamlining the global fit process for the growing ecosystem of HEP Python packages.

\subsection{Results and Discussions}\label{sec: table-results}

The dimension-6 operator map outlining the one-loop matching and running contributions of the scalar and fermion linear SM extensions is presented in Fig.~\ref{fig:map}. Entries denoted by L and R indicate contributions from one-loop matching and running respectively. We also show the tree-level generated operators, marked by a T, for completeness. The colour dimension shows the magnitude of the operator coefficients for unit exotic couplings. Each entry is a link to the dictionary of explicit expressions for the matching results, hosted on GitHub~\href{https://github.com/johngarg/linear-one-loop-dict-ref/tree/main}{\color{red}{\faGithub}}. Additionally, each particle links to an additional directory containing data including the corresponding \texttt{Matchmaker} output and the Python class output by \texttt{MatchMakerParser}~\href{https://github.com/johngarg/MatchMakerParser}{\color{red}{\faGithub}}.

A few patterns are readily apparent. We see that the operators $\mathcal{O}_{G}$, $\mathcal{O}_{W}$, $\mathcal{O}_{HG}$, $\mathcal{O}_{HW}$, $\mathcal{O}_{HB}$ and $\mathcal{O}_{HWB}$ can only be generated by loop-level matching, with no contributions from tree-level matching or one-loop running, while $\mathcal{O}_{uG}$, $\mathcal{O}_{uW}$, $\mathcal{O}_{uB}$, $\mathcal{O}_{dG}$, $\mathcal{O}_{dW}$, $\mathcal{O}_{dB}$, $\mathcal{O}_{eW}$, and $\mathcal{O}_{eB}$ can be generated by both one-loop matching and running, but not at tree level. Similar phenomena have been pointed out by Ref.~\cite{Craig:2019wmo}: operators belonging to the classes $F^3, F^2\phi^2, F\psi^2\phi$ are generated at loop level, but only the class $F\psi^2\phi$ is renormalised by tree-level operators, $\psi^4$ (see also Ref.~\cite{Cheung:2015aba} for explanations in terms of helicity amplitudes). This is consistent with the anomalous dimension matrix for the one-loop running of dimension-6 operator coefficients~\cite{Jenkins:2013zja, Alonso:2013hga, Jenkins:2013wua}; for example, from Eq. A19 of \cite{Jenkins:2013wua}, the running of $\mathcal{O}_{eW}$ can be triggered if $\mathcal{O}_{lequ}^{(3)}$ is generated at tree level, so $\mathcal{O}_{eW}$ in the $\Pi_7$ and $\omega_1$ models can be renormalised while in the remaining models it does not run.

All other $CP$-even operators can obtain tree-level contributions from at least one linear SM extension. However, the $CP$-odd operators $\mathcal{O}_{\widetilde{G}}$, $\mathcal{O}_{\widetilde{W}}$, $\mathcal{O}_{H\widetilde{G}}$,
$\mathcal{O}_{H\widetilde{W}}$,
$\mathcal{O}_{H\widetilde{B}}$, and $\mathcal{O}_{HW\widetilde{B}}$, defined in Ref.~\cite{Grzadkowski:2010es}, are not generated either at tree-level or one loop by linear SM extensions. For $\mathcal{O}_{H\widetilde{G}},\mathcal{O}_{H\widetilde{W}},\mathcal{O}_{H\widetilde{B}},\mathcal{O}_{HW\widetilde{B}}$, this is a consequence of our restriction to single-field extensions as those operators can be generated in the case of SM extensions involving at least two fermion fields~\cite{Bakshi:2021ofj}. But for the $CP$-odd triple-gauge operators, $\mathcal{O}_{\widetilde{G}}$ and $\mathcal{O}_{\widetilde{W}}$\,, it can be shown that this does not arise at the one-loop level, see e.g.~Ref.~\cite{Bakshi:2021ofj} that consider specific models and study their Feynman diagrams. A more general demonstration for this fact is succinctly provided by the Universal One-Loop Effective Action (UOLEA)~\cite{Drozd:2015rsp,Ellis:2017jns,Ellis:2020ivx}, where we see that no such universal structure arises regardless of the specific UV model. The $CP$-odd triple-gauge operators must therefore be generated at the 2-loop level. 

Fig.~\ref{fig:map} enables the phenomenology of a particular operator or simplified model to be easily assessed at a glance. Looking down each column, we see the operators relevant for each model of interest. Reading across row-wise tells us which models contribute to a given operator that may enter in a measurement. For operators that can only be generated at loop level, we would miss the implications of their constraints when working only with the tree-level dictionary. Together with a map from observables to operators, as given for example in Ref.~\cite{Castro:2022zpq}, this table streamlines the process of going from theory to data and vice versa and can be valuable in assessing potential anomalies in data (see for example the recent SMEFT analyses of $M_W$~\cite{Bagnaschi:2022whn} and $h \to \gamma\gamma, \gamma Z$~\cite{Mantzaropoulos:2024vpe} that use the tree-level dictionary of linear SM extensions). We illustrate this is in the next Section for $Z$ pole measurements at future colliders. 

\section{Linear SM extensions at a Tera-$Z$ factory}\label{sec:teraZ}

\begin{figure}[t!]
\centering
\includegraphics[scale=0.71]{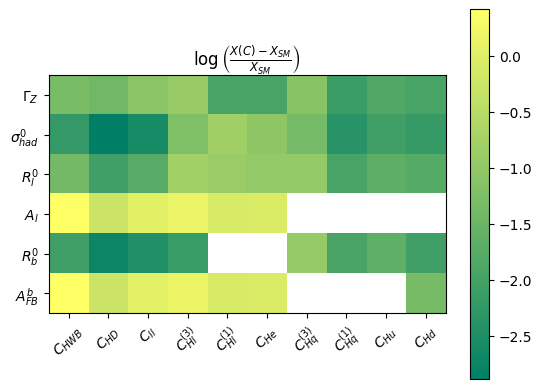}
\caption{Log of the linear dependencies on the Wilson coefficients for each observable in our set of $Z$ pole EWPOs.}
\label{fig:loglineardependencies}
\end{figure}

The relevance of one-loop effects for probing linear SM extensions is especially pertinent at a potential Tera-$Z$ factory, such as the FCC-ee envisioned at CERN~\cite{Bernardi:2022hny} or the CEPC proposed in China~\cite{CEPCStudyGroup:2023quu}. These future $e^+ e^-$ colliders would provide an unprecedented statistical sample of $4-5 \times 10^{12}$ $Z$ bosons, 5 orders of magnitude more than at LEP. Our map streamlines the process of obtaining sensitivity projections and facilitates identification of the relevant one-loop phenomenology, thus serving to demonstrate the importance for probing heavy BSM of a future programme of ultra-high electroweak precision at the $Z$ pole~\cite{Allwicher:2023shc, Stefanek:2024kds, Allwicher:2024sso} that is complementary to new physics probes off the $Z$ pole~\cite{Ge:2024pfn, Greljo:2024ytg}.

The following set of Electroweak Precision Observables (EWPOs) at the $Z$ pole will be used for our analysis: 
\begin{equation}
    \left\{ \Gamma_Z, \, \sigma^0_\text{had}, \, R_l^0 \, , A_l \, , R_b^0 \, , A_{FB}^b \right\} \, .
\end{equation} 
They are defined e.g.~in Refs.~\cite{Wells:2014pga, Allwicher:2024sso} and have been constrained up to the per mille level at LEP which for the most part remain the most precise EWPO measurements to date. There are 10 operators contributing to these EWPOs at leading order in the SMEFT, 
\begin{equation}
    \left\{ \mathcal{O}_{HWB} , \, \mathcal{O}_{HD} , \, \mathcal{O}_{ll} , \, \mathcal{O}^{(3)}_{Hl} , \, \mathcal{O}^{(1)}_{Hl} , \, \mathcal{O}_{He} , \, \mathcal{O}^{(3)}_{Hq} , \, \mathcal{O}^{(1)}_{Hq} , \, \mathcal{O}_{Hu} , \, \mathcal{O}_{Hd} \right\} \, .
    \label{eq:EWPOoperators}
\end{equation}
The linear dependencies of the EWPOs on the Wilson coefficients of these dimension-6 operators are implemented in {\tt Fitmaker}~\cite{Ellis:2020unq}. The log of the linear dependencies of the EWPOs on the SMEFT coefficients are colour coded in the table of Fig.~\ref{fig:loglineardependencies}, indicating which coefficients contribute the most to which observable in the fit. Fig.~\ref{fig:map} can then be used to assess which of the linear SM extensions contribute to these dimension-6 operators at tree or one-loop level. For convenience, the subset of operators relevant for EWPOs are tabulated in Table~\ref{tab:ewpo-op-couplings} with the couplings that enter in the matching listed in each entry. (See Appendix~\ref{sec:Table-couplings} and Table~\ref{tab:ewpo-op-couplings} for the full table of couplings relevant for matching to all other operators.) The entries with rectangular boxes denote tree-level matching while the rest arise only at one loop. We see that at tree level only the following 16 models can be constrained, 
\begin{equation}
\left\{ S_1 , \, \Xi, \, \Xi_1 \, , N \, , E , \, \Delta_1 , \, \Delta_3 , \, \Sigma , \, \Sigma_1 , \, U , \, D , \, Q_1 , \, Q_5 , \, Q_7 , \, T_1 , \, T_2  \right\} \, . 
\end{equation}
On the other hand, there are a further 16 models that have no tree-level contributions to EWPOs but enter at loop level, namely, 
\begin{equation}
\left\{ S , \, S_2 , \, \varphi , \, \Theta_1 , \, \Theta_3 , \, \omega_1 , \, \omega_2 , \, \omega_4 , \, \Pi_1 , \, \Pi_7 , \, \xi , \, \Omega_1 , \, \Omega_2 , \, \Omega_4 , \, \Upsilon , \, \Phi \right\}  \, . 
\end{equation}
A sufficiently high precision to access one-loop effects from heavy new physics is therefore essential to probe the full set of scalar and fermion linear SM extensions in EWPOs. Even in the case of linear SM extensions with tree-level contributions to EWPOs, the couplings responsible for the tree-level matching will necessarily also generate other operators at loop level which may significantly impact the phenomenology if those other operators are more precisely constrained. Should a future pattern of SM deviations be detected, our table would help point towards correlated sets of effects in observables expected under the minimal starting assumption of a single linear SM extension being responsible.

\begin{table}[t!]
\centering
\begin{tabular}{|c|c|c|c|c|c|c|}
\hline
 & $\Gamma_Z$ [GeV] & $\sigma^0_\text{had}$ [nb] & $R_l^0$ & $A_l$ & $R_b^0$ & $A_{FB}^b$ \\
\hline
SM value & 2.4943 & 41.488 & 20.752 & 0.147 & 0.2158  & 0.1031  \\
\hline
Stat. $\sigma$ & $4\times 10^{-6}$ & $1\times 10^{-4}$ & $6 \times 10^{-5}$ & $7\times 10^{-6}$ & $3\times 10^{-7}$  & $2\times 10^{-6}$  \\
\hline
Sys. $\sigma$ & $25\times 10^{-6}$ & $4\times 10^{-3}$ & $1\times 10^{-3}$  & $2\times 10^{-5}$  & $6\times 10^{-5}$  & $1.3\times 10^{-4}$  \\
\hline
Th. $\sigma$ & $150\times 10^{-6}$ & - & $1.5\times 10^{-3}$  & -  & $5\times 10^{-5}$  & - \\
\hline
\end{tabular}
\caption{Projected statistical, systematic and theory uncertainties for $Z$ pole EWPOs at FCC-ee~\cite{Bernardi:2022hny, Freitas:2019bre}.}
\label{tab:EWPOFCC}
\end{table}

\begin{figure}[t!]
\centering
\includegraphics[scale=0.8]{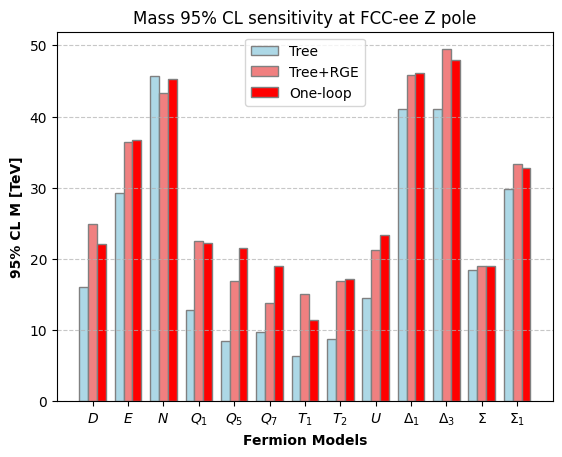}
\caption{Projected 95\% CL sensitivity from FCC-ee Z-pole EWPOs on the mass of fermionic linear SM extensions, matching at tree level (blue bars) and one loop (red bars) to dimension-6 SMEFT operators. The salmon-colored bars are using tree-level plus one-loop log matching contributions only.  }
\label{fig:bar_plot_fermions}
\end{figure}

\begin{figure}[t!]
\centering
\includegraphics[scale=0.8]{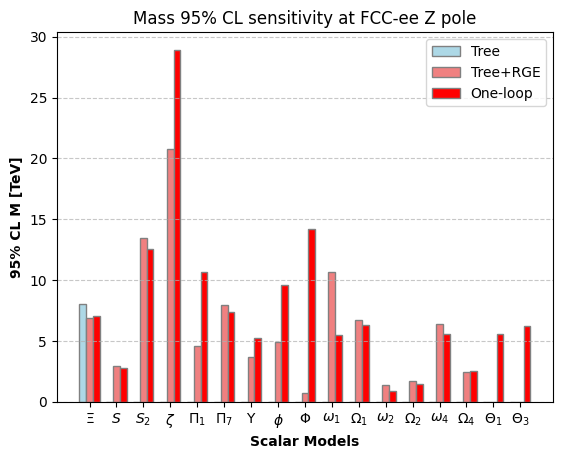}
\caption{Projected 95\% CL sensitivity from FCC-ee Z-pole EWPOs on the mass of scalar linear SM extensions, matching at tree level (blue bars) and one loop (red bars) to dimension-6 SMEFT operators. The salmon-colored bars are using tree-level plus one-loop log matching contributions only.}
\label{fig:bar_plot_scalars}
\end{figure}

The expected statistical, experimental and theoretical systematic uncertainties used in our set of $Z$ pole EWPOs at FCC-ee, taken from Refs.~\cite{Bernardi:2022hny, Freitas:2019bre}, are listed in Table~\ref{tab:EWPOFCC}. We use {\tt Fitmaker}~\cite{Ellis:2020unq} to perform our fit of the scalar and fermion linear SM extensions. All couplings are set to 1. The resulting projected 95\% CL sensitivity to the mass scale in TeV are shown in Figs.~\ref{fig:bar_plot_fermions} and ~\ref{fig:bar_plot_scalars} for the fermions and scalars, respectively. The blue bars denote the fit results using only the matching expressions at tree level, while the red bars use the full one loop matching including finite and logarithmic pieces with the matching scale set to the mass of the field. The salmon-coloured bars are the fit results with matching expressions keeping the logs but discarding the finite one-loop matching parts. This indicates the relative importance of the latter compared to partial one-loop results obtained by tree-level matching and one-loop renormalisation group running of the dimension-6 operators.

\begin{landscape}
\begin{table}[p]
    \centering
    \tiny
    \begin{tabular}{lcccccccccc}
        \toprule
         & $\mathcal{O}_{HWB}$ & $\mathcal{O}_{HD}$ & $\mathcal{O}_{ll}$ & $\mathcal{O}_{Hl}^{(3)}$ & $\mathcal{O}_{Hl}^{(1)}$ & $\mathcal{O}_{He}$ & $\mathcal{O}_{Hq}^{(3)}$ & $\mathcal{O}_{Hq}^{(1)}$ & $\mathcal{O}_{Hu}$ & $\mathcal{O}_{Hd}$ \\
        \midrule
        $S$ & $\kappa_{\mathcal{S}}$ & $\kappa_{\mathcal{S}}$ &  & $\kappa_{\mathcal{S}}$ & $\kappa_{\mathcal{S}}$ & $\kappa_{\mathcal{S}}$ & $\kappa_{\mathcal{S}}$ & $\kappa_{\mathcal{S}}$ & $\kappa_{\mathcal{S}}$ & $\kappa_{\mathcal{S}}$ \\
        $S_1$ &  &  & $\boxed{y_{\mathcal{S}_1}}$ & $y_{\mathcal{S}_1}$ & $y_{\mathcal{S}_1}$ & $y_{\mathcal{S}_1}$ &  &  &  &  \\
        $S_2$ &  &  &  & $y_{\mathcal{S}_2}$ & $y_{\mathcal{S}_2}$ & $y_{\mathcal{S}_2}$ &  &  &  &  \\
        $\varphi$ & $\hat{\lambda}^{\prime}_{\varphi}$ & $\hat{\lambda}^{\prime}_{\varphi}$ & $y_{\varphi e}$ & $y_{\varphi e}$ & $y_{\varphi e}$ & $y_{\varphi e}$ & $y_{\varphi d}$, $y_{\varphi u}$ & $y_{\varphi d}$, $y_{\varphi u}$ & $y_{\varphi d}$, $y_{\varphi u}$ & $y_{\varphi d}$, $y_{\varphi u}$ \\
        $\Xi$ &  & $\boxed{\kappa_{\Xi}}$, $\lambda_{\Xi}$ &  & $\kappa_{\Xi}$ & $\kappa_{\Xi}$ & $\kappa_{\Xi}$ & $\kappa_{\Xi}$ & $\kappa_{\Xi}$ & $\kappa_{\Xi}$ & $\kappa_{\Xi}$ \\
        \multirow{2}{*}{$\Xi_1$} & $\kappa_{\Xi_1}$, ${\lambda_{\Xi_1}^\prime}$ & $\boxed{\kappa_{\Xi_1}}$, $\lambda_{\Xi_1}$ & $\boxed{y_{\Xi_1}}$ & $\kappa_{\Xi_1}$, $y_{\Xi_1}$ & $\kappa_{\Xi_1}$, $y_{\Xi_1}$ & $\kappa_{\Xi_1}$, $y_{\Xi_1}$ & $\kappa_{\Xi_1}$ & $\kappa_{\Xi_1}$ & $\kappa_{\Xi_1}$ & $\kappa_{\Xi_1}$ \\
         &  & ${\lambda_{\Xi_1}^\prime}$ &  &  &  &  &  &  &  &  \\
        \multirow{2}{*}{$\Theta_1$} & $\hat{\lambda}^{\prime}_{\Theta_1}$ & $\hat{\lambda}^{\prime\prime}_{\Theta_1}$, $\hat{\lambda}^{\prime}_{\Theta_1}$ &  &  &  &  &  &  &  &  \\
         &  & $\lambda_{\Theta_1}$ &  &  &  &  &  &  &  &  \\
        $\Theta_3$ & $\hat{\lambda}^{\prime}_{\Theta_3}$ & $\hat{\lambda}^{\prime}_{\Theta_3}$, $\lambda_{\Theta_3}$ &  &  &  &  &  &  &  &  \\
        \multirow{2}{*}{$\omega_1$} &  &  & $y_{q\ell \omega_1}$ & $y_{e u \omega_1}$, $y_{q\ell \omega_1}$ & $y_{e u \omega_1}$, $y_{q\ell \omega_1}$ & $y_{e u \omega_1}$, $y_{q\ell \omega_1}$ & $y_{d u \omega_1}$, $y_{e u \omega_1}$ & $y_{d u \omega_1}$, $y_{e u \omega_1}$ & $y_{d u \omega_1}$, $y_{e u \omega_1}$ & $y_{d u \omega_1}$, $y_{q\ell \omega_1}$ \\
         &  &  &  &  &  &  & $y_{q\ell \omega_1}$, $y_{qq \omega_1}$ & $y_{q\ell \omega_1}$, $y_{qq \omega_1}$ & $y_{q\ell \omega_1}$, $y_{qq \omega_1}$ & $y_{qq \omega_1}$ \\
        $\omega_2$ &  &  &  &  &  &  & $y_{\omega_2}$ & $y_{\omega_2}$ &  & $y_{\omega_2}$ \\
        $\omega_4$ &  &  &  & $y_{e d \omega_4}$ & $y_{e d \omega_4}$ & $y_{e d \omega_4}$ & $y_{e d \omega_4}$, $y_{u u \omega_4}$ & $y_{e d \omega_4}$, $y_{u u \omega_4}$ & $y_{u u \omega_4}$ & $y_{e d \omega_4}$ \\
        $\Pi_1$ & $\hat{\lambda}^{\prime}_{\Pi_1}$ & $\hat{\lambda}^{\prime}_{\Pi_1}$ & $y_{\Pi_1}$ & $y_{\Pi_1}$ & $y_{\Pi_1}$ & $y_{\Pi_1}$ & $y_{\Pi_1}$ & $y_{\Pi_1}$ &  & $y_{\Pi_1}$ \\
        $\Pi_7$ & $\hat{\lambda}^{\prime}_{\Pi_7}$ & $\hat{\lambda}^{\prime}_{\Pi_7}$ & $y_{\ell u \Pi_7}$ & $y_{e q \Pi_7}$, $y_{\ell u \Pi_7}$ & $y_{e q \Pi_7}$, $y_{\ell u \Pi_7}$ & $y_{e q \Pi_7}$, $y_{\ell u \Pi_7}$ & $y_{e q \Pi_7}$, $y_{\ell u \Pi_7}$ & $y_{e q \Pi_7}$, $y_{\ell u \Pi_7}$ & $y_{e q \Pi_7}$, $y_{\ell u \Pi_7}$ & $y_{e q \Pi_7}$ \\
        $\zeta$ & $\hat{\lambda}^{\prime}_{\zeta}$ & $\hat{\lambda}^{\prime}_{\zeta}$ & $y_{q\ell \zeta}$ & $y_{q\ell \zeta}$ & $y_{q\ell \zeta}$ & $y_{q\ell \zeta}$ & $y_{q\ell \zeta}$, $y_{qq \zeta}$ & $y_{q\ell \zeta}$, $y_{qq \zeta}$ & $y_{q\ell \zeta}$, $y_{qq \zeta}$ & $y_{q\ell \zeta}$, $y_{qq \zeta}$ \\
        $\Omega_1$ &  &  &  &  &  &  & $y_{q q \Omega_1}$, $y_{u d \Omega_1}$ & $y_{q q \Omega_1}$, $y_{u d \Omega_1}$ & $y_{q q \Omega_1}$, $y_{u d \Omega_1}$ & $y_{q q \Omega_1}$, $y_{u d \Omega_1}$ \\
        $\Omega_2$ &  &  &  &  &  &  & $y_{\Omega_2}$ & $y_{\Omega_2}$ &  & $y_{\Omega_2}$ \\
        $\Omega_4$ &  &  &  &  &  &  & $y_{\Omega_4}$ & $y_{\Omega_4}$ & $y_{\Omega_4}$ &  \\
        $\Upsilon$ & $\hat{\lambda}^{\prime}_{\Upsilon}$ & $\hat{\lambda}^{\prime}_{\Upsilon}$ &  &  &  &  & $y_{\Upsilon}$ & $y_{\Upsilon}$ & $y_{\Upsilon}$ & $y_{\Upsilon}$ \\
        $\Phi$ & $\hat{\lambda}^{\prime}_{\Phi}$ & $\hat{\lambda}^{\prime}_{\Phi}$, $\hat{\lambda}^{\prime\prime}_{\Phi}$ &  &  &  &  & $y_{q d \Phi}$, $y_{q u \Phi}$ & $y_{q d \Phi}$, $y_{q u \Phi}$ & $y_{q d \Phi}$, $y_{q u \Phi}$ & $y_{q d \Phi}$, $y_{q u \Phi}$ \\
        $N$ & $\lambda_N$ & $\lambda_N$ & $\lambda_N$ & $\boxed{\lambda_N}$ & $\boxed{\lambda_N}$ & $\lambda_N$ & $\lambda_N$ & $\lambda_N$ & $\lambda_N$ & $\lambda_N$ \\
        $E$ & $\lambda_E$ & $\lambda_E$ & $\lambda_E$ & $\boxed{\lambda_E}$ & $\boxed{\lambda_E}$ & $\lambda_E$ & $\lambda_E$ & $\lambda_E$ & $\lambda_E$ & $\lambda_E$ \\
        $\Delta_1$ & $\lambda_{\Delta_1}$ & $\lambda_{\Delta_1}$ &  & $\lambda_{\Delta_1}$ & $\lambda_{\Delta_1}$ & $\boxed{\lambda_{\Delta_1}}$ & $\lambda_{\Delta_1}$ & $\lambda_{\Delta_1}$ & $\lambda_{\Delta_1}$ & $\lambda_{\Delta_1}$ \\
        $\Delta_3$ & $\lambda_{\Delta_3}$ & $\lambda_{\Delta_3}$ &  & $\lambda_{\Delta_3}$ & $\lambda_{\Delta_3}$ & $\boxed{\lambda_{\Delta_3}}$ & $\lambda_{\Delta_3}$ & $\lambda_{\Delta_3}$ & $\lambda_{\Delta_3}$ & $\lambda_{\Delta_3}$ \\
        $\Sigma$ & $\lambda_{\Sigma}$ & $\lambda_{\Sigma}$ & $\lambda_{\Sigma}$ & $\boxed{\lambda_{\Sigma}}$ & $\boxed{\lambda_{\Sigma}}$ & $\lambda_{\Sigma}$ & $\lambda_{\Sigma}$ & $\lambda_{\Sigma}$ & $\lambda_{\Sigma}$ & $\lambda_{\Sigma}$ \\
        $\Sigma_1$ & $\lambda_{\Sigma_1}$ & $\lambda_{\Sigma_1}$ & $\lambda_{\Sigma_1}$ & $\boxed{\lambda_{\Sigma_1}}$ & $\boxed{\lambda_{\Sigma_1}}$ & $\lambda_{\Sigma_1}$ & $\lambda_{\Sigma_1}$ & $\lambda_{\Sigma_1}$ & $\lambda_{\Sigma_1}$ & $\lambda_{\Sigma_1}$ \\
        $U$ & $\lambda_U$ & $\lambda_U$ &  & $\lambda_U$ & $\lambda_U$ & $\lambda_U$ & $\boxed{\lambda_U}$ & $\boxed{\lambda_U}$ & $\lambda_U$ & $\lambda_U$ \\
        $D$ &  & $\lambda_D$ &  & $\lambda_D$ & $\lambda_D$ & $\lambda_D$ & $\boxed{\lambda_D}$ & $\boxed{\lambda_D}$ & $\lambda_D$ & $\lambda_D$ \\
        $Q_1$ & $\lambda_{d Q_1}$, $\lambda_{u Q_1}$ & $\lambda_{d Q_1}$, $\lambda_{u Q_1}$ &  & $\lambda_{d Q_1}$, $\lambda_{u Q_1}$ & $\lambda_{d Q_1}$, $\lambda_{u Q_1}$ & $\lambda_{d Q_1}$, $\lambda_{u Q_1}$ & $\lambda_{d Q_1}$, $\lambda_{u Q_1}$ & $\lambda_{d Q_1}$, $\lambda_{u Q_1}$ & $\lambda_{d Q_1}$, $\boxed{\lambda_{u Q_1}}$ & $\boxed{\lambda_{d Q_1}}$, $\lambda_{u Q_1}$ \\
        $Q_5$ & $\lambda_{Q_5}$ & $\lambda_{Q_5}$ &  & $\lambda_{Q_5}$ & $\lambda_{Q_5}$ & $\lambda_{Q_5}$ & $\lambda_{Q_5}$ & $\lambda_{Q_5}$ & $\lambda_{Q_5}$ & $\boxed{\lambda_{Q_5}}$ \\
        $Q_7$ & $\lambda_{Q_7}$ & $\lambda_{Q_7}$ &  & $\lambda_{Q_7}$ & $\lambda_{Q_7}$ & $\lambda_{Q_7}$ & $\lambda_{Q_7}$ & $\lambda_{Q_7}$ & $\boxed{\lambda_{Q_7}}$ & $\lambda_{Q_7}$ \\
        $T_1$ & $\lambda_{T_1}$ & $\lambda_{T_1}$ &  & $\lambda_{T_1}$ & $\lambda_{T_1}$ & $\lambda_{T_1}$ & $\boxed{\lambda_{T_1}}$ & $\boxed{\lambda_{T_1}}$ & $\lambda_{T_1}$ & $\lambda_{T_1}$ \\
        $T_2$ & $\lambda_{T_2}$ & $\lambda_{T_2}$ &  & $\lambda_{T_2}$ & $\lambda_{T_2}$ & $\lambda_{T_2}$ & $\boxed{\lambda_{T_2}}$ & $\boxed{\lambda_{T_2}}$ & $\lambda_{T_2}$ & $\lambda_{T_2}$ \\
        \bottomrule
        \end{tabular}
  \caption{\label{tab:ewpo-op-couplings}
   The table shows the exotic couplings appearing in the matching expressions for the operators shown. Coupling constants appearing at tree level are shown boxed. Flavour indices have been suppressed.}
\end{table}
\end{landscape}

For the fermion models, we see in Fig.~\ref{fig:bar_plot_fermions} a change in sensitivity in most cases when using the full one-loop matching results driven by the additional contributions generated by the finite one-loop matching. For example, while both the $S$ and $T$ parameters, corresponding to the operators $\mathcal{O}_{HWB}$ and $\mathcal{O}_{HD}$ respectively, are generated exclusively at one loop, the $S$ parameter does not receive any log contributions; one-loop RGE running cannot induce a contribution to $\mathcal{O}_{HWB}$ since the operators that enter the anomalous dimension matrix, $\mathcal{O}_W$, $\mathcal{O}_{HW}$, and $\mathcal{O}_{HB}$, can themselves only be generated at one-loop level by the linear SM extensions. These features and more are evident from Fig.~\ref{fig:map}. We find that all fermion linear SM extensions can be probed for masses at least above 10 TeV and up to about 50 TeV, setting the couplings to 1. 

For the scalar models, the one-loop contributions are even more crucial to properly assess their sensitivity as the vast majority do not enter at tree level in EWPOs. The finite one-loop effects can lead to a significant difference compared with only including their RGE-generated contributions, as in the case of $\zeta, \Phi, \Theta_1$ and $\Theta_3$. The small log term for $\Phi$ is due to a cancellation between the $y_{qd\Phi}$ and $y_{du\Phi}$ contributions; setting one of them to zero increases the sensitivity to 3.6 TeV. $\Theta_1$ and $\Theta_3$ on the other hand receive no RGE-induced contributions into EWPOs at one loop from their tree-level matched operator, since only $\mathcal{O}_H$ is generated at tree level and this does not run into any of the operators in~\ref{eq:EWPOoperators}. The improvement in the constraints when including finite one-loop contributions come not only from one-loop-generated operators due to the tree-level couplings, but also from additional couplings that are relevant for one-loop matching but do not lead to tree-level matching. These effects are all included in our analysis. For example, we see that $\Phi$ gets a much stronger constraint from finite one-loop contributions to the $S$ and $T$ parameters that are generated by couplings that only enter at one loop, as denoted by the hats on the couplings in Table~\ref{tab:ewpo-op-couplings}.

In general, Fig.~\ref{fig:bar_plot_scalars} shows that FCC-ee can reach a sensitivity ranging from a few TeV to around 30 TeV from the scalar one loop contributions even with unit couplings. Only two models, the scalars $S_1$ and $\Xi_1$, contribute at tree level to the highly sensitive four-fermion operator $\mathcal{O}_{ll}$.\footnote{We note that due to the antisymmetry of the $S_1$ couplings to the different generations it should vanish when coupling to identical fields.} $\Xi_1$ also generates $\mathcal{O}_{HD}$ at tree level which enters with an equal but opposite contribution to $\mathcal{O}_{ll}$ in the SMEFT input parameter shifts, so there is an exact cancellation at leading order when $y_{\Xi_1} = \kappa_{\Xi_1} = 1$. The tree-level sensitivity to $S_1$ and $\Xi_1$ (assuming $\kappa_{\Xi_1} = 0)$ from $\mathcal{O}_{ll}$ is $\sim 75$ TeV. The loop level contributions are omitted in this case due to large logs from such a high scale that would then have to be resummed.  

In this analysis we have made some simplifying assumptions, such as fixing all couplings to 1. The projected reach to the mass scale of the linear SM extensions could increase or decrease depending on whether they are strongly or weakly coupled, respectively. We have also assumed one model at a time in keeping with our simplified model approach; more realistic UV-complete scenarios may involve several degrees of freedom appearing at a time in the same observables. However, it is clear from our analysis that relatively generic new physics at the TeV scale, if they exist and couple linearly to the SM, will likely show indirect effects in EWPOs at FCC-ee.

\section{Conclusion}\label{sec:Conclusion}

The phenomenology of linear SM extensions is becoming an increasingly important benchmark to assess the sensitivity of direct and indirect searches. This finite set of simplified models is motivated by being the only possibilities for BSM fields coupling linearly to the SM with relevant or marginal interactions, which enables them to be singly produced at colliders for direct searches and to match at tree-level onto the SMEFT for indirect searches.  

We extended the available tree-level dictionary of dimension-6 operators for the scalar and fermion linear SM extensions to include the one-loop matching results. Such a systematic and extensive computation has only recently become feasible with the advent of improved automation tools and matching techniques that are still in active development~\cite{Fuentes-Martin:2023ljp,Born:2024mgz,Li:2024ciy}. Our application here, summarised in Fig.~\ref{fig:map}, not only enables the analytic expressions to be readily accessible through clickable links for each table entry and for the particles names, it also provides a high-level overview at a glance of the relevant phenomenology for each model and operator of interest. This will serve as a valuable reference for future SMEFT studies that wish to isolate phenomena due to linear SM extensions as a minimal benchmark.    

As an illustration of the utility of our results, we studied the projected sensitivity to the linear SM extensions of a future $Z$ pole run at the FCC-ee proposed by CERN. Tera-$Z$ statistics enable ultra-high precision measurements, limited only by systematic and theory uncertainties, that will be indirectly sensitive to heavy new physics at the TeV scale even for weakly coupled new physics generated at one loop. We find that a Tera-$Z$ factory will probe all the scalar and fermion linear SM extensions up to almost 50 TeV for unit couplings. This highlights the importance of a future programme of electroweak precision measurements for a thorough quantum exploration of the high-energy frontier. 

Directions for future studies include the possibility of including 2 linear SM extensions at a time for each operator. Such an exercise would reveal any further so-called magic zeroes in one-loop matching that have been identified in certain cases~\cite{Arkani-Hamed:2021xlp,Craig:2021ksw,DelleRose:2022ygn,Guedes:2022cfy}, or new enhancement for the SM couplings (e.g. including two fermionic extensions can universally enhance all light quark Yukawa couplings, see Ref.~\cite{Erdelyi:2024sls}). This exercise will provide a further step towards mapping the phenomenology of these simplified models onto more realistic UV theories where more than one field typically appears at a time. Our results could be interfaced to a global fit in the SMEFT at current and future colliders for a more complete picture of indirect constraints on heavy linear SM extensions. It would also be interesting to consider the complementarity between indirect and direct searches for the linear SM extension particles at future high-energy colliders such as a muon collider or FCC-hh. 

\section*{Acknowledgments}
\sloppy % Necessary to prevent the grant numbers running into the margin
We thank Lukas Allwicher, Guilherme Guedes, Victor Maura, Pablo Olgoso, Jos\'e Santiago, Ben Stefanek, and Anders Eller Thomsen for useful discussions. JG was partially funded by the Spanish ``Agencia Estatal de Investigaci\'on'' MICIN/AEI/10.13039/501100011033 through the grant PID2020-113334GB-I00 and by the ``Juan de la Cierva'' programme reference FJC2021-048111-I, both also funded by ``European Union NextGenerationEU/PRTR''. JG is also supported by the ARC Centre of Excellence for Dark Matter Particle Physics CE20010000.~ 
\unskip % Close the `\sloppy` above 
The work of PNHV is supported by the Deutsche Forschungsgemeinschaft (DFG, German Research Foundation) under Germany's Excellence Strategy -- EXC 2121 ``Quantum Universe" -- 390833306, as well as by the grant 491245950. TY is supported by United Kingdom Science and Technologies Facilities Council (STFC) grant ST/X000753/1.

\appendix

\section{Code repositories}
\label{sec:repositories}

We have made the code associated with our analysis and our results available through three publicly available GitHub repositories. Below we summarise these and provide the links in one place for easy reference.

\begin{enumerate}
\item \texttt{MatchMakerParser}~\href{https://github.com/johngarg/MatchMakerParser}{\color{red}{\faGithub}} --- This is the tool that parses the symbolic expressions from Mathematica code to Python code. The reader need not interact with this if they are only concerned with the one-loop results relevant to the linear SM extensions.
\item \texttt{linear-one-loop-dict-ref}~\href{https://github.com/johngarg/linear-one-loop-dict-ref}{\color{red}{\faGithub}} --- This repository hosts the data linked to in Fig.~\ref{fig:map}. It is only intended to be used in conjunction with our main table of results where the individual results are more easily accessible by clicking the interactive entries in Fig.~\ref{fig:map}.
\item \texttt{lsme}~\href{https://github.com/johngarg/lsme}{\color{red}{\faGithub}} --- This page hosts an up-to-date copy of the code published as the Python package \texttt{lsme}. This package is the most convenient Pythonic interface to our one-loop dictionary. The \texttt{readme} file contains information on how to install and use the package. 
\end{enumerate}

\section{Tables of couplings}
\label{sec:Table-couplings}

Below we present Tables~\ref{tab:couplings-table-1}--\ref{tab:couplings-table-4}, which show the exotic coupling constants appearing in the matching expressions for each model and operator. The exotic couplings appearing in the matching expressions for the operators contributing to electroweak precision observables are shown in Table~\ref{tab:ewpo-op-couplings}, and for this reason those operators are omitted below. Couplings appearing already through tree-level matching are shown boxed, and all flavour indices have been suppressed in the tables. We note that in some cases operator coefficients depend on only SM couplings constants. This is true for one-loop contributions to the triple-gauge operators $\mathcal{O}_{G}$ and $\mathcal{O}_{W}$ in the models we study, and so we leave these off the listings below. The absence of one-loop contributions to $CP$-odd bosonic operators in the models we study allows us to omit these from the tables below as well. See Sec.~\ref{sec:One-loop map of SMEFT} for a more detailed discussion of these operators and the topologies that generate them.

\begin{landscape}
  \begin{table}[t]
    \centering
    \tiny

    \begin{tabular}{lcccccccccc}
\toprule
 & $\mathcal{O}_{H}$ & $\mathcal{O}_{HB}$ & $\mathcal{O}_{H D^2}$ & $\mathcal{O}_{HG}$ & $\mathcal{O}_{HW}$ & $\mathcal{O}_{Hud}$ & $\mathcal{O}_{dB}$ & $\mathcal{O}_{dG}$ & $\mathcal{O}_{dH}$ & $\mathcal{O}_{dW}$ \\
\midrule
\multirow{2}{*}{$S$} & $\boxed{\kappa_{\mathcal{S}}}$, $\boxed{\kappa_{\mathcal{S}3}}$ & $\kappa_{\mathcal{S}}$ & $\boxed{\kappa_{\mathcal{S}}}$, $\kappa_{\mathcal{S}3}$ &  & $\kappa_{\mathcal{S}}$ & $\kappa_{\mathcal{S}}$ &  &  & $\kappa_{\mathcal{S}}$, $\kappa_{\mathcal{S}3}$ &  \\
 & $\boxed{\lambda_{\mathcal{S}}}$ &  & $\lambda_{\mathcal{S}}$ &  &  &  &  &  & $\lambda_{\mathcal{S}}$ &  \\
$S_1$ & $\hat{\lambda}_{\mathcal{S}_1}$ & $\hat{\lambda}_{\mathcal{S}_1}$ & $\hat{\lambda}_{\mathcal{S}_1}$ &  &  &  &  &  &  &  \\
$S_2$ & $\hat{\lambda}_{\mathcal{S}_2}$ & $\hat{\lambda}_{\mathcal{S}_2}$ & $\hat{\lambda}_{\mathcal{S}_2}$ &  &  &  &  &  &  &  \\
\multirow{2}{*}{$\varphi$} & $\hat{\lambda}^{\prime}_{\varphi}$, $\hat{\lambda}_{\varphi}$ & $\hat{\lambda}^{\prime}_{\varphi}$, $\hat{\lambda}_{\varphi}$ & $\hat{\lambda}^{\prime}_{\varphi}$, $\hat{\lambda}_{\varphi}$ &  & $\hat{\lambda}^{\prime}_{\varphi}$, $\hat{\lambda}_{\varphi}$ & $y_{\varphi d}$, $y_{\varphi u}$ & $y_{\varphi d}$, $y_{\varphi u}$ & $y_{\varphi d}$, $y_{\varphi u}$ & $\hat{\lambda}^{\prime}_{\varphi}$, $\hat{\lambda}_{\varphi}$ & $y_{\varphi d}$, $y_{\varphi u}$ \\
 & $\boxed{\lambda_{\varphi}}$ &  & $\lambda_{\varphi}$ &  &  &  &  &  & $\boxed{\lambda_{\varphi}}$, $\boxed{y_{\varphi d}}$, $y_{\varphi u}$ &  \\
$\Xi$ & $\boxed{\kappa_{\Xi}}$, $\boxed{\lambda_{\Xi}}$ & $\kappa_{\Xi}$ & $\boxed{\kappa_{\Xi}}$, $\lambda_{\Xi}$ &  & $\kappa_{\Xi}$, $\lambda_{\Xi}$ & $\kappa_{\Xi}$ &  &  & $\boxed{\kappa_{\Xi}}$, $\lambda_{\Xi}$ &  \\
\multirow{2}{*}{$\Xi_1$} & $\boxed{\kappa_{\Xi_1}}$, $\boxed{\lambda_{\Xi_1}}$ & $\kappa_{\Xi_1}$, $\lambda_{\Xi_1}$ & $\boxed{\kappa_{\Xi_1}}$, $\lambda_{\Xi_1}$ &  & $\kappa_{\Xi_1}$, $\lambda_{\Xi_1}$ &  &  &  & $\boxed{\kappa_{\Xi_1}}$, $\lambda_{\Xi_1}$ &  \\
 & $\boxed{{\lambda_{\Xi_1}^\prime}}$ &  & ${\lambda_{\Xi_1}^\prime}$ &  &  &  &  &  & ${\lambda_{\Xi_1}^\prime}$ &  \\
\multirow{2}{*}{$\Theta_1$} & $\hat{\lambda}^{\prime\prime}_{\Theta_1}$, $\hat{\lambda}^{\prime}_{\Theta_1}$ & $\hat{\lambda}_{\Theta_1}$ & $\hat{\lambda}^{\prime\prime}_{\Theta_1}$, $\hat{\lambda}^{\prime}_{\Theta_1}$ &  & $\hat{\lambda}_{\Theta_1}$ &  &  &  & $\hat{\lambda}^{\prime\prime}_{\Theta_1}$, $\hat{\lambda}^{\prime}_{\Theta_1}$ &  \\
 & $\hat{\lambda}_{\Theta_1}$, $\boxed{\lambda_{\Theta_1}}$ &  & $\hat{\lambda}_{\Theta_1}$, $\lambda_{\Theta_1}$ &  &  &  &  &  & $\lambda_{\Theta_1}$ &  \\
\multirow{2}{*}{$\Theta_3$} & $\hat{\lambda}^{\prime}_{\Theta_3}$, $\hat{\lambda}_{\Theta_3}$ & $\hat{\lambda}_{\Theta_3}$ & $\hat{\lambda}^{\prime}_{\Theta_3}$, $\hat{\lambda}_{\Theta_3}$ &  & $\hat{\lambda}_{\Theta_3}$ &  &  &  & $\hat{\lambda}^{\prime}_{\Theta_3}$, $\lambda_{\Theta_3}$ &  \\
 & $\boxed{\lambda_{\Theta_3}}$ &  & $\lambda_{\Theta_3}$ &  &  &  &  &  &  &  \\
\multirow{2}{*}{$\omega_1$} & $\hat{\lambda}_{\omega_1}$ & $\hat{\lambda}_{\omega_1}$ & $\hat{\lambda}_{\omega_1}$ & $\hat{\lambda}_{\omega_1}$ &  & $y_{d u \omega_1}$, $y_{e u \omega_1}$ & $y_{d u \omega_1}$, $y_{q\ell \omega_1}$ & $y_{d u \omega_1}$, $y_{q\ell \omega_1}$ & $\hat{\lambda}_{\omega_1}$, $y_{d u \omega_1}$ & $y_{d u \omega_1}$, $y_{q\ell \omega_1}$ \\
 &  &  &  &  &  & $y_{q\ell \omega_1}$, $y_{qq \omega_1}$ & $y_{qq \omega_1}$ & $y_{qq \omega_1}$ & $y_{q\ell \omega_1}$, $y_{qq \omega_1}$ & $y_{qq \omega_1}$ \\
$\omega_2$ & $\hat{\lambda}_{\omega_2}$ & $\hat{\lambda}_{\omega_2}$ & $\hat{\lambda}_{\omega_2}$ & $\hat{\lambda}_{\omega_2}$ &  &  & $y_{\omega_2}$ & $y_{\omega_2}$ & $\hat{\lambda}_{\omega_2}$, $y_{\omega_2}$ &  \\
$\omega_4$ & $\hat{\lambda}_{\omega_4}$ & $\hat{\lambda}_{\omega_4}$ & $\hat{\lambda}_{\omega_4}$ & $\hat{\lambda}_{\omega_4}$ &  &  & $y_{e d \omega_4}$ & $y_{e d \omega_4}$ & $\hat{\lambda}_{\omega_4}$, $y_{e d \omega_4}$ &  \\
\multirow{2}{*}{$\Pi_1$} & $\hat{\lambda}_{\Pi_1}$, $\hat{\lambda}^{\prime}_{\Pi_1}$ & $\hat{\lambda}_{\Pi_1}$, $\hat{\lambda}^{\prime}_{\Pi_1}$ & $\hat{\lambda}_{\Pi_1}$, $\hat{\lambda}^{\prime}_{\Pi_1}$ & $\hat{\lambda}_{\Pi_1}$, $\hat{\lambda}^{\prime}_{\Pi_1}$ & $\hat{\lambda}_{\Pi_1}$, $\hat{\lambda}^{\prime}_{\Pi_1}$ &  & $y_{\Pi_1}$ & $y_{\Pi_1}$ & $\hat{\lambda}_{\Pi_1}$, $\hat{\lambda}^{\prime}_{\Pi_1}$ &  \\
 &  &  &  &  &  &  &  &  & $y_{\Pi_1}$ &  \\
\multirow{2}{*}{$\Pi_7$} & $\hat{\lambda}_{\Pi_7}$, $\hat{\lambda}^{\prime}_{\Pi_7}$ & $\hat{\lambda}_{\Pi_7}$, $\hat{\lambda}^{\prime}_{\Pi_7}$ & $\hat{\lambda}_{\Pi_7}$, $\hat{\lambda}^{\prime}_{\Pi_7}$ & $\hat{\lambda}_{\Pi_7}$, $\hat{\lambda}^{\prime}_{\Pi_7}$ & $\hat{\lambda}_{\Pi_7}$, $\hat{\lambda}^{\prime}_{\Pi_7}$ & $y_{e q \Pi_7}$, $y_{\ell u \Pi_7}$ & $y_{e q \Pi_7}$ & $y_{e q \Pi_7}$ & $\hat{\lambda}_{\Pi_7}$, $\hat{\lambda}^{\prime}_{\Pi_7}$ & $y_{e q \Pi_7}$ \\
 &  &  &  &  &  &  &  &  & $y_{e q \Pi_7}$ &  \\
\multirow{2}{*}{$\zeta$} & $\hat{\lambda}^{\prime}_{\zeta}$, $\hat{\lambda}_{\zeta}$ & $\hat{\lambda}_{\zeta}$ & $\hat{\lambda}^{\prime}_{\zeta}$, $\hat{\lambda}_{\zeta}$ & $\hat{\lambda}_{\zeta}$ & $\hat{\lambda}_{\zeta}$ & $y_{q\ell \zeta}$, $y_{qq \zeta}$ & $y_{q\ell \zeta}$, $y_{qq \zeta}$ & $y_{q\ell \zeta}$, $y_{qq \zeta}$ & $\hat{\lambda}^{\prime}_{\zeta}$, $\hat{\lambda}_{\zeta}$ & $y_{q\ell \zeta}$, $y_{qq \zeta}$ \\
 &  &  &  &  &  &  &  &  & $y_{q\ell \zeta}$, $y_{qq \zeta}$ &  \\
\multirow{2}{*}{$\Omega_1$} & $\hat{\lambda}_{\Omega_1}$ & $\hat{\lambda}_{\Omega_1}$ & $\hat{\lambda}_{\Omega_1}$ & $\hat{\lambda}_{\Omega_1}$ &  & $y_{q q \Omega_1}$, $y_{u d \Omega_1}$ & $y_{q q \Omega_1}$, $y_{u d \Omega_1}$ & $y_{q q \Omega_1}$, $y_{u d \Omega_1}$ & $\hat{\lambda}_{\Omega_1}$, $y_{q q \Omega_1}$ & $y_{q q \Omega_1}$, $y_{u d \Omega_1}$ \\
 &  &  &  &  &  &  &  &  & $y_{u d \Omega_1}$ &  \\
$\Omega_2$ & $\hat{\lambda}_{\Omega_2}$ & $\hat{\lambda}_{\Omega_2}$ & $\hat{\lambda}_{\Omega_2}$ & $\hat{\lambda}_{\Omega_2}$ &  &  & $y_{\Omega_2}$ & $y_{\Omega_2}$ & $\hat{\lambda}_{\Omega_2}$, $y_{\Omega_2}$ &  \\
$\Omega_4$ & $\hat{\lambda}_{\Omega_4}$ & $\hat{\lambda}_{\Omega_4}$ & $\hat{\lambda}_{\Omega_4}$ & $\hat{\lambda}_{\Omega_4}$ &  &  &  &  &  &  \\
\multirow{2}{*}{$\Upsilon$} & $\hat{\lambda}^{\prime}_{\Upsilon}$, $\hat{\lambda}_{\Upsilon}$ & $\hat{\lambda}_{\Upsilon}$ & $\hat{\lambda}^{\prime}_{\Upsilon}$, $\hat{\lambda}_{\Upsilon}$ & $\hat{\lambda}_{\Upsilon}$ & $\hat{\lambda}_{\Upsilon}$ & $y_{\Upsilon}$ & $y_{\Upsilon}$ & $y_{\Upsilon}$ & $\hat{\lambda}^{\prime}_{\Upsilon}$, $\hat{\lambda}_{\Upsilon}$ & $y_{\Upsilon}$ \\
 &  &  &  &  &  &  &  &  & $y_{\Upsilon}$ &  \\
\multirow{2}{*}{$\Phi$} & $\hat{\lambda}_{\Phi}$, $\hat{\lambda}^{\prime}_{\Phi}$ & $\hat{\lambda}_{\Phi}$, $\hat{\lambda}^{\prime}_{\Phi}$ & $\hat{\lambda}_{\Phi}$, $\hat{\lambda}^{\prime}_{\Phi}$ & $\hat{\lambda}_{\Phi}$, $\hat{\lambda}^{\prime}_{\Phi}$ & $\hat{\lambda}_{\Phi}$, $\hat{\lambda}^{\prime}_{\Phi}$ & $y_{q d \Phi}$, $y_{q u \Phi}$ & $y_{q d \Phi}$, $y_{q u \Phi}$ & $y_{q d \Phi}$, $y_{q u \Phi}$ & $\hat{\lambda}_{\Phi}$, $\hat{\lambda}^{\prime}_{\Phi}$ & $y_{q d \Phi}$, $y_{q u \Phi}$ \\
 & $\hat{\lambda}^{\prime\prime}_{\Phi}$ &  & $\hat{\lambda}^{\prime\prime}_{\Phi}$ &  &  &  &  &  & $\hat{\lambda}^{\prime\prime}_{\Phi}$, $y_{q d \Phi}$, $y_{q u \Phi}$ &  \\
$N$ & $\lambda_N$ & $\lambda_N$ & $\lambda_N$ &  & $\lambda_N$ &  &  &  & $\lambda_N$ &  \\
$E$ & $\lambda_E$ & $\lambda_E$ & $\lambda_E$ &  & $\lambda_E$ &  &  &  & $\lambda_E$ &  \\
$\Delta_1$ & $\lambda_{\Delta_1}$ & $\lambda_{\Delta_1}$ & $\lambda_{\Delta_1}$ &  &  &  &  &  & $\lambda_{\Delta_1}$ &  \\
$\Delta_3$ & $\lambda_{\Delta_3}$ & $\lambda_{\Delta_3}$ & $\lambda_{\Delta_3}$ &  &  &  &  &  & $\lambda_{\Delta_3}$ &  \\
$\Sigma$ & $\lambda_{\Sigma}$ & $\lambda_{\Sigma}$ & $\lambda_{\Sigma}$ &  & $\lambda_{\Sigma}$ &  &  &  & $\lambda_{\Sigma}$ &  \\
$\Sigma_1$ & $\lambda_{\Sigma_1}$ & $\lambda_{\Sigma_1}$ & $\lambda_{\Sigma_1}$ &  & $\lambda_{\Sigma_1}$ &  &  &  & $\lambda_{\Sigma_1}$ &  \\
$U$ & $\lambda_U$ & $\lambda_U$ & $\lambda_U$ & $\lambda_U$ & $\lambda_U$ & $\lambda_U$ & $\lambda_U$ & $\lambda_U$ & $\lambda_U$ & $\lambda_U$ \\
$D$ & $\lambda_D$ & $\lambda_D$ & $\lambda_D$ & $\lambda_D$ & $\lambda_D$ & $\lambda_D$ & $\lambda_D$ & $\lambda_D$ & $\boxed{\lambda_D}$ & $\lambda_D$ \\
$Q_1$ & $\lambda_{d Q_1}$, $\lambda_{u Q_1}$ & $\lambda_{d Q_1}$, $\lambda_{u Q_1}$ & $\lambda_{d Q_1}$, $\lambda_{u Q_1}$ & $\lambda_{d Q_1}$, $\lambda_{u Q_1}$ &  & $\boxed{\lambda_{d Q_1}}$, $\boxed{\lambda_{u Q_1}}$ & $\lambda_{d Q_1}$, $\lambda_{u Q_1}$ & $\lambda_{d Q_1}$, $\lambda_{u Q_1}$ & $\boxed{\lambda_{d Q_1}}$, $\lambda_{u Q_1}$ & $\lambda_{d Q_1}$, $\lambda_{u Q_1}$ \\
$Q_5$ & $\lambda_{Q_5}$ & $\lambda_{Q_5}$ & $\lambda_{Q_5}$ & $\lambda_{Q_5}$ &  & $\lambda_{Q_5}$ & $\lambda_{Q_5}$ & $\lambda_{Q_5}$ & $\boxed{\lambda_{Q_5}}$ & $\lambda_{Q_5}$ \\
$Q_7$ & $\lambda_{Q_7}$ & $\lambda_{Q_7}$ & $\lambda_{Q_7}$ & $\lambda_{Q_7}$ &  & $\lambda_{Q_7}$ &  &  & $\lambda_{Q_7}$ &  \\
$T_1$ & $\lambda_{T_1}$ & $\lambda_{T_1}$ & $\lambda_{T_1}$ & $\lambda_{T_1}$ & $\lambda_{T_1}$ & $\lambda_{T_1}$ & $\lambda_{T_1}$ & $\lambda_{T_1}$ & $\boxed{\lambda_{T_1}}$ & $\lambda_{T_1}$ \\
$T_2$ & $\lambda_{T_2}$ & $\lambda_{T_2}$ & $\lambda_{T_2}$ & $\lambda_{T_2}$ & $\lambda_{T_2}$ & $\lambda_{T_2}$ & $\lambda_{T_2}$ & $\lambda_{T_2}$ & $\boxed{\lambda_{T_2}}$ & $\lambda_{T_2}$ \\
\bottomrule
\end{tabular}

    \caption{\label{tab:couplings-table-1}
       The table shows the exotic couplings appearing in the matching expressions for the operators shown. Coupling constants appearing at tree level are shown boxed. Flavour indices have been suppressed.}
  \end{table}
\end{landscape}

\begin{landscape}
  \begin{table}[t]
    \centering
    \tiny
    \begin{tabular}{lcccccccccccc}
\toprule
 & $\mathcal{O}_{dd}$ & $\mathcal{O}_{eB}$ & $\mathcal{O}_{eH}$ & $\mathcal{O}_{eW}$ & $\mathcal{O}_{ed}$ & $\mathcal{O}_{ee}$ & $\mathcal{O}_{eu}$ & $\mathcal{O}_{ld}$ & $\mathcal{O}_{le}$ & $\mathcal{O}_{ledq}$ & $\mathcal{O}_{lequ}^{(1)}$ & $\mathcal{O}_{lequ}^{(3)}$ \\
\midrule
\multirow{2}{*}{$S$} &  &  & $\kappa_{\mathcal{S}}$, $\kappa_{\mathcal{S}3}$ &  &  &  &  &  & $\kappa_{\mathcal{S}}$ & $\kappa_{\mathcal{S}}$ & $\kappa_{\mathcal{S}}$ &  \\
 &  &  & $\lambda_{\mathcal{S}}$ &  &  &  &  &  &  &  &  &  \\
$S_1$ &  & $y_{\mathcal{S}_1}$ & $\hat{\lambda}_{\mathcal{S}_1}$, $y_{\mathcal{S}_1}$ & $y_{\mathcal{S}_1}$ &  &  &  & $y_{\mathcal{S}_1}$ & $y_{\mathcal{S}_1}$ &  &  &  \\
$S_2$ &  & $y_{\mathcal{S}_2}$ & $\hat{\lambda}_{\mathcal{S}_2}$, $y_{\mathcal{S}_2}$ &  & $y_{\mathcal{S}_2}$ & $\boxed{y_{\mathcal{S}_2}}$ & $y_{\mathcal{S}_2}$ &  & $y_{\mathcal{S}_2}$ &  &  &  \\
\multirow{2}{*}{$\varphi$} & $y_{\varphi d}$ & $y_{\varphi e}$ & $\hat{\lambda}^{\prime}_{\varphi}$, $\hat{\lambda}_{\varphi}$ & $y_{\varphi e}$ & $y_{\varphi d}$, $y_{\varphi e}$ & $y_{\varphi e}$ & $y_{\varphi e}$, $y_{\varphi u}$ & $y_{\varphi d}$, $y_{\varphi e}$ & $\boxed{y_{\varphi e}}$ & $\boxed{y_{\varphi d}}$, $\boxed{y_{\varphi e}}$ & $y_{\varphi d}$, $\boxed{y_{\varphi e}}$ & $y_{\varphi e}$, $y_{\varphi u}$ \\
 &  &  & $\boxed{\lambda_{\varphi}}$, $\boxed{y_{\varphi e}}$ &  &  &  &  &  &  & $y_{\varphi u}$ & $\boxed{y_{\varphi u}}$ &  \\
$\Xi$ &  &  & $\boxed{\kappa_{\Xi}}$, $\lambda_{\Xi}$ &  &  &  &  &  & $\kappa_{\Xi}$ & $\kappa_{\Xi}$ & $\kappa_{\Xi}$ &  \\
\multirow{2}{*}{$\Xi_1$} &  & $y_{\Xi_1}$ & $\boxed{\kappa_{\Xi_1}}$, $\lambda_{\Xi_1}$ & $y_{\Xi_1}$ &  &  &  & $y_{\Xi_1}$ & $\kappa_{\Xi_1}$, $y_{\Xi_1}$ & $\kappa_{\Xi_1}$ & $\kappa_{\Xi_1}$ &  \\
 &  &  & ${\lambda_{\Xi_1}^\prime}$, $y_{\Xi_1}$ &  &  &  &  &  &  &  &  &  \\
\multirow{2}{*}{$\Theta_1$} &  &  & $\hat{\lambda}^{\prime\prime}_{\Theta_1}$, $\hat{\lambda}^{\prime}_{\Theta_1}$ &  &  &  &  &  &  &  &  &  \\
 &  &  & $\lambda_{\Theta_1}$ &  &  &  &  &  &  &  &  &  \\
$\Theta_3$ &  &  & $\hat{\lambda}^{\prime}_{\Theta_3}$, $\lambda_{\Theta_3}$ &  &  &  &  &  &  &  &  &  \\
\multirow{2}{*}{$\omega_1$} & $y_{d u \omega_1}$ & $y_{e u \omega_1}$, $y_{q\ell \omega_1}$ & $\hat{\lambda}_{\omega_1}$, $y_{e u \omega_1}$ & $y_{e u \omega_1}$, $y_{q\ell \omega_1}$ & $y_{d u \omega_1}$, $y_{e u \omega_1}$ & $y_{e u \omega_1}$ & $y_{d u \omega_1}$, $\boxed{y_{e u \omega_1}}$ & $y_{d u \omega_1}$, $y_{q\ell \omega_1}$ & $y_{e u \omega_1}$, $y_{q\ell \omega_1}$ & $y_{d u \omega_1}$, $y_{e u \omega_1}$ & $y_{d u \omega_1}$, $\boxed{y_{e u \omega_1}}$ & $y_{d u \omega_1}$, $\boxed{y_{e u \omega_1}}$ \\
 &  &  & $y_{q\ell \omega_1}$ &  &  &  & $y_{q\ell \omega_1}$ &  &  & $y_{q\ell \omega_1}$, $y_{qq \omega_1}$ & $\boxed{y_{q\ell \omega_1}}$, $y_{qq \omega_1}$ & $\boxed{y_{q\ell \omega_1}}$, $y_{qq \omega_1}$ \\
$\omega_2$ & $\boxed{y_{\omega_2}}$ &  &  &  & $y_{\omega_2}$ &  &  & $y_{\omega_2}$ &  &  &  &  \\
$\omega_4$ & $y_{e d \omega_4}$ & $y_{e d \omega_4}$ & $\hat{\lambda}_{\omega_4}$, $y_{e d \omega_4}$ &  & $\boxed{y_{e d \omega_4}}$ & $y_{e d \omega_4}$ & $y_{e d \omega_4}$, $y_{u u \omega_4}$ & $y_{e d \omega_4}$ & $y_{e d \omega_4}$ & $y_{e d \omega_4}$ &  &  \\
\multirow{2}{*}{$\Pi_1$} & $y_{\Pi_1}$ & $y_{\Pi_1}$ & $\hat{\lambda}_{\Pi_1}$, $\hat{\lambda}^{\prime}_{\Pi_1}$ & $y_{\Pi_1}$ & $y_{\Pi_1}$ &  &  & $\boxed{y_{\Pi_1}}$ & $y_{\Pi_1}$ & $y_{\Pi_1}$ &  &  \\
 &  &  & $y_{\Pi_1}$ &  &  &  &  &  &  &  &  &  \\
\multirow{2}{*}{$\Pi_7$} &  & $y_{e q \Pi_7}$, $y_{\ell u \Pi_7}$ & $\hat{\lambda}_{\Pi_7}$, $\hat{\lambda}^{\prime}_{\Pi_7}$ & $y_{e q \Pi_7}$, $y_{\ell u \Pi_7}$ & $y_{e q \Pi_7}$ & $y_{e q \Pi_7}$ & $y_{e q \Pi_7}$, $y_{\ell u \Pi_7}$ & $y_{\ell u \Pi_7}$ & $y_{e q \Pi_7}$, $y_{\ell u \Pi_7}$ & $y_{e q \Pi_7}$, $y_{\ell u \Pi_7}$ & $\boxed{y_{e q \Pi_7}}$, $\boxed{y_{\ell u \Pi_7}}$ & $\boxed{y_{e q \Pi_7}}$, $\boxed{y_{\ell u \Pi_7}}$ \\
 &  &  & $y_{e q \Pi_7}$, $y_{\ell u \Pi_7}$ &  &  &  &  &  &  &  &  &  \\
\multirow{2}{*}{$\zeta$} &  &  & $\hat{\lambda}^{\prime}_{\zeta}$, $\hat{\lambda}_{\zeta}$ & $y_{q\ell \zeta}$ &  &  &  & $y_{q\ell \zeta}$ & $y_{q\ell \zeta}$ & $y_{q\ell \zeta}$ &  &  \\
 &  &  & $y_{q\ell \zeta}$ &  &  &  &  &  &  &  &  &  \\
$\Omega_1$ & $y_{u d \Omega_1}$ &  &  &  & $y_{u d \Omega_1}$ &  & $y_{u d \Omega_1}$ & $y_{u d \Omega_1}$ &  & $y_{q q \Omega_1}$, $y_{u d \Omega_1}$ & $y_{q q \Omega_1}$, $y_{u d \Omega_1}$ &  \\
$\Omega_2$ & $\boxed{y_{\Omega_2}}$ &  &  &  & $y_{\Omega_2}$ &  &  & $y_{\Omega_2}$ &  &  &  &  \\
$\Omega_4$ &  &  &  &  &  &  & $y_{\Omega_4}$ &  &  &  &  &  \\
$\Upsilon$ &  &  & $\hat{\lambda}^{\prime}_{\Upsilon}$ &  &  &  &  &  &  &  &  &  \\
$\Phi$ & $y_{q d \Phi}$ &  & $\hat{\lambda}^{\prime}_{\Phi}$, $\hat{\lambda}^{\prime\prime}_{\Phi}$ &  & $y_{q d \Phi}$ &  & $y_{q u \Phi}$ & $y_{q d \Phi}$ &  & $y_{q d \Phi}$, $y_{q u \Phi}$ & $y_{q d \Phi}$, $y_{q u \Phi}$ &  \\
$N$ &  & $\lambda_N$ & $\lambda_N$ & $\lambda_N$ &  &  &  & $\lambda_N$ & $\lambda_N$ & $\lambda_N$ & $\lambda_N$ &  \\
$E$ &  & $\lambda_E$ & $\boxed{\lambda_E}$ & $\lambda_E$ &  &  &  & $\lambda_E$ & $\lambda_E$ & $\lambda_E$ & $\lambda_E$ &  \\
$\Delta_1$ &  & $\lambda_{\Delta_1}$ & $\boxed{\lambda_{\Delta_1}}$ &  & $\lambda_{\Delta_1}$ & $\lambda_{\Delta_1}$ & $\lambda_{\Delta_1}$ &  & $\lambda_{\Delta_1}$ & $\lambda_{\Delta_1}$ & $\lambda_{\Delta_1}$ &  \\
$\Delta_3$ &  & $\lambda_{\Delta_3}$ & $\boxed{\lambda_{\Delta_3}}$ & $\lambda_{\Delta_3}$ & $\lambda_{\Delta_3}$ & $\lambda_{\Delta_3}$ & $\lambda_{\Delta_3}$ &  & $\lambda_{\Delta_3}$ & $\lambda_{\Delta_3}$ & $\lambda_{\Delta_3}$ &  \\
$\Sigma$ &  & $\lambda_{\Sigma}$ & $\boxed{\lambda_{\Sigma}}$ & $\lambda_{\Sigma}$ &  &  &  & $\lambda_{\Sigma}$ & $\lambda_{\Sigma}$ & $\lambda_{\Sigma}$ & $\lambda_{\Sigma}$ &  \\
$\Sigma_1$ &  & $\lambda_{\Sigma_1}$ & $\boxed{\lambda_{\Sigma_1}}$ & $\lambda_{\Sigma_1}$ &  &  &  & $\lambda_{\Sigma_1}$ & $\lambda_{\Sigma_1}$ & $\lambda_{\Sigma_1}$ & $\lambda_{\Sigma_1}$ &  \\
$U$ &  &  & $\lambda_U$ &  &  &  &  &  & $\lambda_U$ & $\lambda_U$ & $\lambda_U$ &  \\
$D$ &  &  & $\lambda_D$ &  &  &  &  &  & $\lambda_D$ & $\lambda_D$ & $\lambda_D$ &  \\
$Q_1$ & $\lambda_{d Q_1}$ &  & $\lambda_{d Q_1}$, $\lambda_{u Q_1}$ &  & $\lambda_{d Q_1}$ &  & $\lambda_{u Q_1}$ & $\lambda_{d Q_1}$ & $\lambda_{d Q_1}$, $\lambda_{u Q_1}$ & $\lambda_{d Q_1}$, $\lambda_{u Q_1}$ & $\lambda_{d Q_1}$, $\lambda_{u Q_1}$ &  \\
$Q_5$ & $\lambda_{Q_5}$ &  & $\lambda_{Q_5}$ &  & $\lambda_{Q_5}$ &  &  & $\lambda_{Q_5}$ & $\lambda_{Q_5}$ & $\lambda_{Q_5}$ & $\lambda_{Q_5}$ &  \\
$Q_7$ &  &  & $\lambda_{Q_7}$ &  &  &  & $\lambda_{Q_7}$ &  & $\lambda_{Q_7}$ & $\lambda_{Q_7}$ & $\lambda_{Q_7}$ &  \\
$T_1$ &  &  & $\lambda_{T_1}$ &  &  &  &  &  & $\lambda_{T_1}$ & $\lambda_{T_1}$ & $\lambda_{T_1}$ &  \\
$T_2$ &  &  & $\lambda_{T_2}$ &  &  &  &  &  & $\lambda_{T_2}$ & $\lambda_{T_2}$ & $\lambda_{T_2}$ &  \\
\bottomrule
\end{tabular}
    \caption{\label{tab:couplings-table-2}
       The table shows the exotic couplings appearing in the matching expressions for the operators shown. Coupling constants appearing at tree level are shown boxed. Flavour indices have been suppressed.}
  \end{table}
\end{landscape}

\begin{landscape}
  \begin{table}[t]
    \centering
    \tiny

\begin{tabular}{lccccccccc}
\toprule
 & $\mathcal{O}_{lq}^{(1)}$ & $\mathcal{O}_{lq}^{(3)}$ & $\mathcal{O}_{lu}$ & $\mathcal{O}_{qd}^{(1)}$ & $\mathcal{O}_{qd}^{(8)}$ & $\mathcal{O}_{qe}$ & $\mathcal{O}_{qq}^{(1)}$ & $\mathcal{O}_{qq}^{(3)}$ & $\mathcal{O}_{qu}^{(1)}$ \\
\midrule
$S$ &  &  &  & $\kappa_{\mathcal{S}}$ & $\kappa_{\mathcal{S}}$ &  &  &  & $\kappa_{\mathcal{S}}$ \\
$S_1$ & $y_{\mathcal{S}_1}$ & $y_{\mathcal{S}_1}$ & $y_{\mathcal{S}_1}$ &  &  &  &  &  &  \\
$S_2$ &  &  &  &  &  & $y_{\mathcal{S}_2}$ &  &  &  \\
\multirow{2}{*}{$\varphi$} & $y_{\varphi d}$, $y_{\varphi e}$ & $y_{\varphi d}$, $y_{\varphi e}$ & $y_{\varphi e}$, $y_{\varphi u}$ & $\boxed{y_{\varphi d}}$, $y_{\varphi u}$ & $\boxed{y_{\varphi d}}$, $y_{\varphi u}$ & $y_{\varphi d}$, $y_{\varphi e}$ & $y_{\varphi d}$, $y_{\varphi u}$ & $y_{\varphi d}$, $y_{\varphi u}$ & $y_{\varphi d}$, $\boxed{y_{\varphi u}}$ \\
 & $y_{\varphi u}$ & $y_{\varphi u}$ &  &  &  & $y_{\varphi u}$ &  &  &  \\
$\Xi$ &  &  &  & $\kappa_{\Xi}$ & $\kappa_{\Xi}$ &  &  &  & $\kappa_{\Xi}$ \\
$\Xi_1$ & $y_{\Xi_1}$ & $y_{\Xi_1}$ & $y_{\Xi_1}$ & $\kappa_{\Xi_1}$ & $\kappa_{\Xi_1}$ &  &  &  & $\kappa_{\Xi_1}$ \\
$\Theta_1$ &  &  &  &  &  &  &  &  &  \\
$\Theta_3$ &  &  &  &  &  &  &  &  &  \\
\multirow{2}{*}{$\omega_1$} & $y_{e u \omega_1}$, $\boxed{y_{q\ell \omega_1}}$ & $y_{e u \omega_1}$, $\boxed{y_{q\ell \omega_1}}$ & $y_{d u \omega_1}$, $y_{e u \omega_1}$ & $y_{d u \omega_1}$, $y_{q\ell \omega_1}$ & $y_{d u \omega_1}$, $y_{q\ell \omega_1}$ & $y_{e u \omega_1}$, $y_{q\ell \omega_1}$ & $y_{d u \omega_1}$, $y_{q\ell \omega_1}$ & $y_{d u \omega_1}$, $y_{q\ell \omega_1}$ & $y_{d u \omega_1}$, $y_{e u \omega_1}$ \\
 & $y_{qq \omega_1}$ & $y_{qq \omega_1}$ & $y_{q\ell \omega_1}$ & $y_{qq \omega_1}$ & $y_{qq \omega_1}$ & $y_{qq \omega_1}$ & $\boxed{y_{qq \omega_1}}$ & $\boxed{y_{qq \omega_1}}$ & $y_{q\ell \omega_1}$, $y_{qq \omega_1}$ \\
$\omega_2$ &  &  &  & $y_{\omega_2}$ & $y_{\omega_2}$ &  &  &  &  \\
$\omega_4$ &  &  & $y_{u u \omega_4}$ & $y_{e d \omega_4}$ & $y_{e d \omega_4}$ & $y_{e d \omega_4}$ &  &  & $y_{u u \omega_4}$ \\
$\Pi_1$ & $y_{\Pi_1}$ & $y_{\Pi_1}$ & $y_{\Pi_1}$ & $y_{\Pi_1}$ & $y_{\Pi_1}$ &  &  &  &  \\
$\Pi_7$ & $y_{e q \Pi_7}$, $y_{\ell u \Pi_7}$ & $y_{e q \Pi_7}$, $y_{\ell u \Pi_7}$ & $y_{e q \Pi_7}$, $\boxed{y_{\ell u \Pi_7}}$ & $y_{e q \Pi_7}$ & $y_{e q \Pi_7}$ & $\boxed{y_{e q \Pi_7}}$, $y_{\ell u \Pi_7}$ & $y_{e q \Pi_7}$ & $y_{e q \Pi_7}$ & $y_{e q \Pi_7}$, $y_{\ell u \Pi_7}$ \\
$\zeta$ & $\boxed{y_{q\ell \zeta}}$, $y_{qq \zeta}$ & $\boxed{y_{q\ell \zeta}}$, $y_{qq \zeta}$ & $y_{q\ell \zeta}$ & $y_{q\ell \zeta}$, $y_{qq \zeta}$ & $y_{q\ell \zeta}$, $y_{qq \zeta}$ & $y_{q\ell \zeta}$, $y_{qq \zeta}$ & $y_{q\ell \zeta}$, $\boxed{y_{qq \zeta}}$ & $y_{q\ell \zeta}$, $\boxed{y_{qq \zeta}}$ & $y_{q\ell \zeta}$, $y_{qq \zeta}$ \\
$\Omega_1$ & $y_{q q \Omega_1}$ & $y_{q q \Omega_1}$ & $y_{u d \Omega_1}$ & $y_{q q \Omega_1}$, $y_{u d \Omega_1}$ & $y_{q q \Omega_1}$, $y_{u d \Omega_1}$ & $y_{q q \Omega_1}$ & $\boxed{y_{q q \Omega_1}}$, $y_{u d \Omega_1}$ & $\boxed{y_{q q \Omega_1}}$, $y_{u d \Omega_1}$ & $y_{q q \Omega_1}$, $y_{u d \Omega_1}$ \\
$\Omega_2$ &  &  &  & $y_{\Omega_2}$ & $y_{\Omega_2}$ &  &  &  &  \\
$\Omega_4$ &  &  & $y_{\Omega_4}$ &  &  &  &  &  & $y_{\Omega_4}$ \\
$\Upsilon$ & $y_{\Upsilon}$ & $y_{\Upsilon}$ &  & $y_{\Upsilon}$ & $y_{\Upsilon}$ & $y_{\Upsilon}$ & $\boxed{y_{\Upsilon}}$ & $\boxed{y_{\Upsilon}}$ & $y_{\Upsilon}$ \\
$\Phi$ & $y_{q d \Phi}$, $y_{q u \Phi}$ & $y_{q d \Phi}$, $y_{q u \Phi}$ & $y_{q u \Phi}$ & $\boxed{y_{q d \Phi}}$, $y_{q u \Phi}$ & $\boxed{y_{q d \Phi}}$, $y_{q u \Phi}$ & $y_{q d \Phi}$, $y_{q u \Phi}$ & $y_{q d \Phi}$, $y_{q u \Phi}$ & $y_{q d \Phi}$, $y_{q u \Phi}$ & $y_{q d \Phi}$, $\boxed{y_{q u \Phi}}$ \\
$N$ & $\lambda_N$ & $\lambda_N$ & $\lambda_N$ & $\lambda_N$ & $\lambda_N$ &  &  &  & $\lambda_N$ \\
$E$ & $\lambda_E$ & $\lambda_E$ & $\lambda_E$ & $\lambda_E$ & $\lambda_E$ &  &  &  & $\lambda_E$ \\
$\Delta_1$ &  &  &  & $\lambda_{\Delta_1}$ & $\lambda_{\Delta_1}$ & $\lambda_{\Delta_1}$ &  &  & $\lambda_{\Delta_1}$ \\
$\Delta_3$ &  &  &  & $\lambda_{\Delta_3}$ & $\lambda_{\Delta_3}$ & $\lambda_{\Delta_3}$ &  &  & $\lambda_{\Delta_3}$ \\
$\Sigma$ & $\lambda_{\Sigma}$ & $\lambda_{\Sigma}$ & $\lambda_{\Sigma}$ & $\lambda_{\Sigma}$ & $\lambda_{\Sigma}$ &  &  &  & $\lambda_{\Sigma}$ \\
$\Sigma_1$ & $\lambda_{\Sigma_1}$ & $\lambda_{\Sigma_1}$ & $\lambda_{\Sigma_1}$ & $\lambda_{\Sigma_1}$ & $\lambda_{\Sigma_1}$ &  &  &  & $\lambda_{\Sigma_1}$ \\
$U$ & $\lambda_U$ & $\lambda_U$ &  & $\lambda_U$ & $\lambda_U$ & $\lambda_U$ & $\lambda_U$ & $\lambda_U$ & $\lambda_U$ \\
$D$ & $\lambda_D$ & $\lambda_D$ &  & $\lambda_D$ & $\lambda_D$ & $\lambda_D$ & $\lambda_D$ & $\lambda_D$ & $\lambda_D$ \\
$Q_1$ &  &  & $\lambda_{u Q_1}$ & $\lambda_{d Q_1}$, $\lambda_{u Q_1}$ & $\lambda_{d Q_1}$, $\lambda_{u Q_1}$ &  &  &  & $\lambda_{d Q_1}$, $\lambda_{u Q_1}$ \\
$Q_5$ &  &  &  & $\lambda_{Q_5}$ & $\lambda_{Q_5}$ &  &  &  & $\lambda_{Q_5}$ \\
$Q_7$ &  &  & $\lambda_{Q_7}$ & $\lambda_{Q_7}$ & $\lambda_{Q_7}$ &  &  &  & $\lambda_{Q_7}$ \\
$T_1$ & $\lambda_{T_1}$ & $\lambda_{T_1}$ &  & $\lambda_{T_1}$ & $\lambda_{T_1}$ & $\lambda_{T_1}$ & $\lambda_{T_1}$ & $\lambda_{T_1}$ & $\lambda_{T_1}$ \\
$T_2$ & $\lambda_{T_2}$ & $\lambda_{T_2}$ &  & $\lambda_{T_2}$ & $\lambda_{T_2}$ & $\lambda_{T_2}$ & $\lambda_{T_2}$ & $\lambda_{T_2}$ & $\lambda_{T_2}$ \\
\bottomrule
\end{tabular}
    
    \caption{\label{tab:couplings-table-3}
       The table shows the exotic couplings appearing in the matching expressions for the operators shown. Coupling constants appearing at tree level are shown boxed. Flavour indices have been suppressed.}
  \end{table}
\end{landscape}

\begin{landscape}
  \begin{table}[t]
    \centering
    \tiny
    \begin{tabular}{lcccccccccc}
\toprule
 & $\mathcal{O}_{qu}^{(8)}$ & $\mathcal{O}_{quqd}^{(1)}$ & $\mathcal{O}_{quqd}^{(8)}$ & $\mathcal{O}_{uB}$ & $\mathcal{O}_{uG}$ & $\mathcal{O}_{uH}$ & $\mathcal{O}_{uW}$ & $\mathcal{O}_{ud}^{(1)}$ & $\mathcal{O}_{ud}^{(8)}$ & $\mathcal{O}_{uu}$ \\
\midrule
\multirow{2}{*}{$S$} & $\kappa_{\mathcal{S}}$ & $\kappa_{\mathcal{S}}$ &  &  &  & $\kappa_{\mathcal{S}}$, $\kappa_{\mathcal{S}3}$ &  &  &  &  \\
 &  &  &  &  &  & $\lambda_{\mathcal{S}}$ &  &  &  &  \\
$S_1$ &  &  &  &  &  &  &  &  &  &  \\
$S_2$ &  &  &  &  &  &  &  &  &  &  \\
\multirow{2}{*}{$\varphi$} & $y_{\varphi d}$, $\boxed{y_{\varphi u}}$ & $\boxed{y_{\varphi d}}$, $\boxed{y_{\varphi u}}$ & $y_{\varphi d}$, $y_{\varphi u}$ & $y_{\varphi d}$, $y_{\varphi u}$ & $y_{\varphi d}$, $y_{\varphi u}$ & $\hat{\lambda}^{\prime}_{\varphi}$, $\hat{\lambda}_{\varphi}$ & $y_{\varphi d}$, $y_{\varphi u}$ & $y_{\varphi d}$, $y_{\varphi u}$ & $y_{\varphi d}$, $y_{\varphi u}$ & $y_{\varphi u}$ \\
 &  &  &  &  &  & $\boxed{\lambda_{\varphi}}$, $y_{\varphi d}$, $\boxed{y_{\varphi u}}$ &  &  &  &  \\
$\Xi$ & $\kappa_{\Xi}$ & $\kappa_{\Xi}$ &  &  &  & $\boxed{\kappa_{\Xi}}$, $\lambda_{\Xi}$ &  &  &  &  \\
\multirow{2}{*}{$\Xi_1$} & $\kappa_{\Xi_1}$ & $\kappa_{\Xi_1}$ &  &  &  & $\boxed{\kappa_{\Xi_1}}$, $\lambda_{\Xi_1}$ &  &  &  &  \\
 &  &  &  &  &  & ${\lambda_{\Xi_1}^\prime}$ &  &  &  &  \\
\multirow{2}{*}{$\Theta_1$} &  &  &  &  &  & $\hat{\lambda}^{\prime\prime}_{\Theta_1}$, $\hat{\lambda}^{\prime}_{\Theta_1}$ &  &  &  &  \\
 &  &  &  &  &  & $\lambda_{\Theta_1}$ &  &  &  &  \\
$\Theta_3$ &  &  &  &  &  & $\hat{\lambda}^{\prime}_{\Theta_3}$, $\lambda_{\Theta_3}$ &  &  &  &  \\
\multirow{2}{*}{$\omega_1$} & $y_{d u \omega_1}$, $y_{e u \omega_1}$ & $\boxed{y_{d u \omega_1}}$, $y_{e u \omega_1}$ & $\boxed{y_{d u \omega_1}}$, $y_{e u \omega_1}$ & $y_{d u \omega_1}$, $y_{e u \omega_1}$ & $y_{d u \omega_1}$, $y_{e u \omega_1}$ & $\hat{\lambda}_{\omega_1}$, $y_{d u \omega_1}$ & $y_{d u \omega_1}$, $y_{e u \omega_1}$ & $\boxed{y_{d u \omega_1}}$, $y_{e u \omega_1}$ & $\boxed{y_{d u \omega_1}}$, $y_{e u \omega_1}$ & $y_{d u \omega_1}$, $y_{e u \omega_1}$ \\
 & $y_{q\ell \omega_1}$, $y_{qq \omega_1}$ & $y_{q\ell \omega_1}$, $\boxed{y_{qq \omega_1}}$ & $y_{q\ell \omega_1}$, $\boxed{y_{qq \omega_1}}$ & $y_{q\ell \omega_1}$, $y_{qq \omega_1}$ & $y_{q\ell \omega_1}$, $y_{qq \omega_1}$ & $y_{e u \omega_1}$, $y_{q\ell \omega_1}$, $y_{qq \omega_1}$ & $y_{q\ell \omega_1}$, $y_{qq \omega_1}$ & $y_{qq \omega_1}$ & $y_{qq \omega_1}$ &  \\
$\omega_2$ &  &  &  &  &  &  &  & $y_{\omega_2}$ & $y_{\omega_2}$ &  \\
$\omega_4$ & $y_{u u \omega_4}$ &  &  & $y_{u u \omega_4}$ & $y_{u u \omega_4}$ & $\hat{\lambda}_{\omega_4}$, $y_{u u \omega_4}$ &  & $y_{e d \omega_4}$, $y_{u u \omega_4}$ & $y_{e d \omega_4}$, $y_{u u \omega_4}$ & $\boxed{y_{u u \omega_4}}$ \\
$\Pi_1$ &  &  &  &  &  & $\hat{\lambda}^{\prime}_{\Pi_1}$ &  & $y_{\Pi_1}$ & $y_{\Pi_1}$ &  \\
\multirow{2}{*}{$\Pi_7$} & $y_{e q \Pi_7}$, $y_{\ell u \Pi_7}$ & $y_{e q \Pi_7}$, $y_{\ell u \Pi_7}$ &  & $y_{e q \Pi_7}$, $y_{\ell u \Pi_7}$ & $y_{e q \Pi_7}$, $y_{\ell u \Pi_7}$ & $\hat{\lambda}_{\Pi_7}$, $\hat{\lambda}^{\prime}_{\Pi_7}$ & $y_{e q \Pi_7}$, $y_{\ell u \Pi_7}$ & $y_{\ell u \Pi_7}$ & $y_{\ell u \Pi_7}$ & $y_{\ell u \Pi_7}$ \\
 &  &  &  &  &  & $y_{e q \Pi_7}$, $y_{\ell u \Pi_7}$ &  &  &  &  \\
\multirow{2}{*}{$\zeta$} & $y_{q\ell \zeta}$, $y_{qq \zeta}$ & $y_{qq \zeta}$ & $y_{qq \zeta}$ & $y_{q\ell \zeta}$, $y_{qq \zeta}$ & $y_{q\ell \zeta}$, $y_{qq \zeta}$ & $\hat{\lambda}^{\prime}_{\zeta}$, $\hat{\lambda}_{\zeta}$ & $y_{q\ell \zeta}$, $y_{qq \zeta}$ &  &  &  \\
 &  &  &  &  &  & $y_{q\ell \zeta}$, $y_{qq \zeta}$ &  &  &  &  \\
\multirow{2}{*}{$\Omega_1$} & $y_{q q \Omega_1}$, $y_{u d \Omega_1}$ & $\boxed{y_{q q \Omega_1}}$, $\boxed{y_{u d \Omega_1}}$ & $\boxed{y_{q q \Omega_1}}$, $\boxed{y_{u d \Omega_1}}$ & $y_{q q \Omega_1}$, $y_{u d \Omega_1}$ & $y_{q q \Omega_1}$, $y_{u d \Omega_1}$ & $\hat{\lambda}_{\Omega_1}$, $y_{q q \Omega_1}$ & $y_{q q \Omega_1}$, $y_{u d \Omega_1}$ & $y_{q q \Omega_1}$, $\boxed{y_{u d \Omega_1}}$ & $y_{q q \Omega_1}$, $\boxed{y_{u d \Omega_1}}$ & $y_{u d \Omega_1}$ \\
 &  &  &  &  &  & $y_{u d \Omega_1}$ &  &  &  &  \\
$\Omega_2$ &  &  &  &  &  &  &  & $y_{\Omega_2}$ & $y_{\Omega_2}$ &  \\
$\Omega_4$ & $y_{\Omega_4}$ &  &  & $y_{\Omega_4}$ & $y_{\Omega_4}$ & $\hat{\lambda}_{\Omega_4}$, $y_{\Omega_4}$ &  & $y_{\Omega_4}$ & $y_{\Omega_4}$ & $\boxed{y_{\Omega_4}}$ \\
\multirow{2}{*}{$\Upsilon$} & $y_{\Upsilon}$ &  &  & $y_{\Upsilon}$ & $y_{\Upsilon}$ & $\hat{\lambda}^{\prime}_{\Upsilon}$, $\hat{\lambda}_{\Upsilon}$ & $y_{\Upsilon}$ &  &  &  \\
 &  &  &  &  &  & $y_{\Upsilon}$ &  &  &  &  \\
\multirow{2}{*}{$\Phi$} & $y_{q d \Phi}$, $\boxed{y_{q u \Phi}}$ & $y_{q d \Phi}$, $y_{q u \Phi}$ & $\boxed{y_{q d \Phi}}$, $\boxed{y_{q u \Phi}}$ & $y_{q d \Phi}$, $y_{q u \Phi}$ & $y_{q d \Phi}$, $y_{q u \Phi}$ & $\hat{\lambda}_{\Phi}$, $\hat{\lambda}^{\prime}_{\Phi}$ & $y_{q d \Phi}$, $y_{q u \Phi}$ & $y_{q d \Phi}$, $y_{q u \Phi}$ & $y_{q d \Phi}$, $y_{q u \Phi}$ & $y_{q u \Phi}$ \\
 &  &  &  &  &  & $\hat{\lambda}^{\prime\prime}_{\Phi}$, $y_{q d \Phi}$, $y_{q u \Phi}$ &  &  &  &  \\
$N$ & $\lambda_N$ & $\lambda_N$ &  &  &  & $\lambda_N$ &  &  &  &  \\
$E$ & $\lambda_E$ & $\lambda_E$ &  &  &  & $\lambda_E$ &  &  &  &  \\
$\Delta_1$ & $\lambda_{\Delta_1}$ & $\lambda_{\Delta_1}$ &  &  &  & $\lambda_{\Delta_1}$ &  &  &  &  \\
$\Delta_3$ & $\lambda_{\Delta_3}$ & $\lambda_{\Delta_3}$ &  &  &  & $\lambda_{\Delta_3}$ &  &  &  &  \\
$\Sigma$ & $\lambda_{\Sigma}$ & $\lambda_{\Sigma}$ &  &  &  & $\lambda_{\Sigma}$ &  &  &  &  \\
$\Sigma_1$ & $\lambda_{\Sigma_1}$ & $\lambda_{\Sigma_1}$ &  &  &  & $\lambda_{\Sigma_1}$ &  &  &  &  \\
$U$ & $\lambda_U$ & $\lambda_U$ &  & $\lambda_U$ & $\lambda_U$ & $\boxed{\lambda_U}$ & $\lambda_U$ &  &  &  \\
$D$ & $\lambda_D$ & $\lambda_D$ &  & $\lambda_D$ & $\lambda_D$ & $\lambda_D$ & $\lambda_D$ &  &  &  \\
$Q_1$ & $\lambda_{d Q_1}$, $\lambda_{u Q_1}$ & $\lambda_{d Q_1}$, $\lambda_{u Q_1}$ &  & $\lambda_{d Q_1}$, $\lambda_{u Q_1}$ & $\lambda_{d Q_1}$, $\lambda_{u Q_1}$ & $\lambda_{d Q_1}$, $\boxed{\lambda_{u Q_1}}$ & $\lambda_{d Q_1}$, $\lambda_{u Q_1}$ & $\lambda_{d Q_1}$, $\lambda_{u Q_1}$ & $\lambda_{d Q_1}$, $\lambda_{u Q_1}$ & $\lambda_{u Q_1}$ \\
$Q_5$ & $\lambda_{Q_5}$ & $\lambda_{Q_5}$ &  &  &  & $\lambda_{Q_5}$ &  & $\lambda_{Q_5}$ & $\lambda_{Q_5}$ &  \\
$Q_7$ & $\lambda_{Q_7}$ & $\lambda_{Q_7}$ &  & $\lambda_{Q_7}$ & $\lambda_{Q_7}$ & $\boxed{\lambda_{Q_7}}$ & $\lambda_{Q_7}$ & $\lambda_{Q_7}$ & $\lambda_{Q_7}$ & $\lambda_{Q_7}$ \\
$T_1$ & $\lambda_{T_1}$ & $\lambda_{T_1}$ &  & $\lambda_{T_1}$ & $\lambda_{T_1}$ & $\boxed{\lambda_{T_1}}$ & $\lambda_{T_1}$ &  &  &  \\
$T_2$ & $\lambda_{T_2}$ & $\lambda_{T_2}$ &  & $\lambda_{T_2}$ & $\lambda_{T_2}$ & $\boxed{\lambda_{T_2}}$ & $\lambda_{T_2}$ &  &  &  \\
\bottomrule
\end{tabular}
    \caption{\label{tab:couplings-table-4}
       The table shows the exotic couplings appearing in the matching expressions for the operators shown. Coupling constants appearing at tree level are shown boxed. Flavour indices have been suppressed.}
  \end{table}
\end{landscape}

\section{Projected Tera-$Z$ sensitivity varying couplings}
\label{sec:Table-varying-couplings}

\begin{figure}[ht!]
\centering
\includegraphics[scale=0.8]{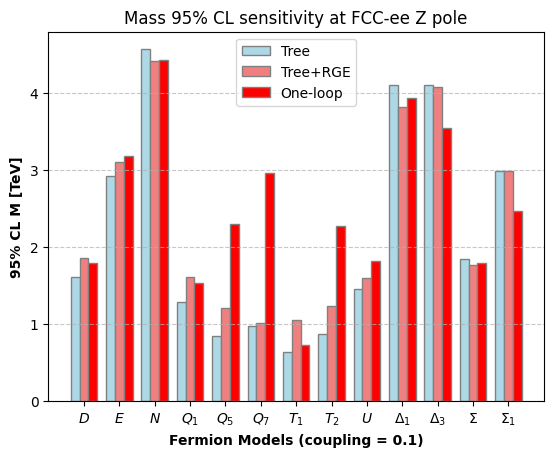}
\includegraphics[scale=0.8]{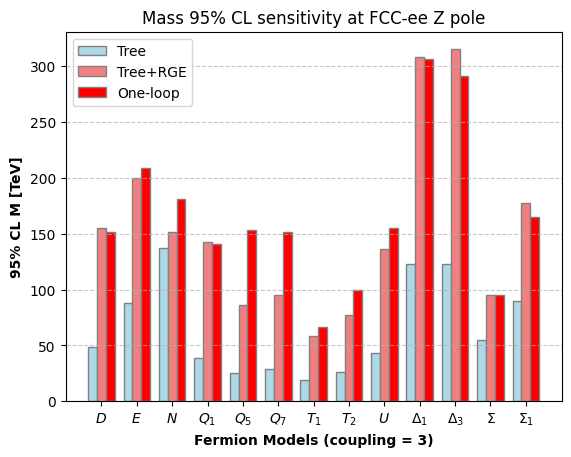}
\caption{Projected 95\% CL sensitivity from FCC-ee Z-pole EWPOs on the mass of fermionic linear SM extensions assuming coupling values of 0.1 (top) and 3 (bottom), matching at tree level (blue bars) and one loop (red bars) to dimension-6 SMEFT operators. The salmon-colored bars are using tree-level plus one-loop log matching contributions only.  }
\label{fig:bar_plot_fermions_app}
\end{figure}

\begin{figure}[t!]
\centering
\includegraphics[scale=0.8]{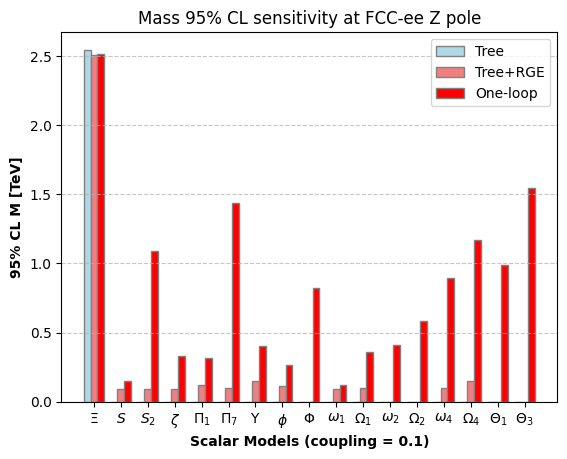}
\includegraphics[scale=0.8]{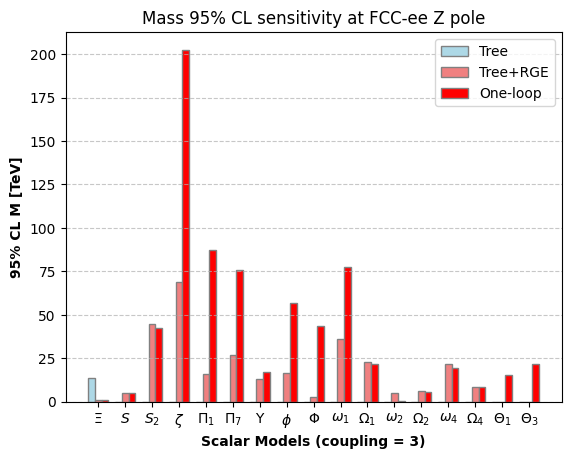}
\caption{Projected 95\% CL sensitivity from FCC-ee Z-pole EWPOs on the mass of scalar linear SM extensions assuming coupling values of 0.1 (top) and 3 (bottom), matching at tree level (blue bars) and one loop (red bars) to dimension-6 SMEFT operators. The salmon-colored bars are using tree-level plus one-loop log matching contributions only.  }
\label{fig:bar_plot_scalars_app}
\end{figure}

We present in Figs.~\ref{fig:bar_plot_fermions_app} and \ref{fig:bar_plot_scalars_app} the projected 95\% CL sensitivities for FCC-ee $Z$-pole EWPOs on the masses of linear SM extensions assuming couplings of 0.1 and 3.0. 

\newpage

\addcontentsline{toc}{section}{References}
\bibliographystyle{JHEP}
\bibliography{main.bib}
\end{document}